# A supCBI process with application to streamflow discharge and a model reduction[a]

Hidekazu Yoshioka[1, *]

[1]Shimane University, Nishikawatsu-cho 1060, Matsue 690-8504, Japan

*Corresponding author: E-mail: yoshih@life.shimane-u.ac.jp

**Abstract**

We propose a new stochastic model for streamflow discharge timeseries as a jump-driven process, called a superposition of continuous-state branching processes with immigration (a supCBI process). It is a non-Markovian model having the capability of reproducing the subexponential autocorrelation found in the hydrological data. The Markovian embedding as a version of matrix analytic methods is applied to the supCBI process, successfully yielding analytical formulae of statistical moments and autocorrelation. The supCBI process is identified at study sites, where hourly streamflow discharge data are available. We also consider another Markovian embedding as a model reduction of the supCBI process to a continuous-time binary semi-Markov chain of high- and low-flow regimes. We show that waiting times can be modeled using a mixture of exponential distributions, suggesting that semi-Markov chains serve as effectively reduced models of the supCBI process.

---

[a] This work was prepared as a manuscript version of a presentation at the conference *The 11th Matrix-Analytic Methods in Stochastic Models*. This preprint is Version 1 on June 13, 2022.

**Introduction**

*Background*

Streamflow is an essential driver shaping the environments and ecosystems of rivers, their watersheds, and their connected surface water bodies, such as lakes and seas (Lapides et al., 2021; Lapides et al., 2022; Saharia et al., 2021). In addition, streamflow directly affects human lives because it controls the availability of water resources (Hwang and Devineni, 2022; Ward et al., 2022) and flood damage (He et al., 2022; Liu et al., 2022). Dams as central water infrastructures are controlled primarily by streamflow dynamics. (Santos et al., 2022; Wang et al., 2022; Yoshioka and Yoshioka, 2020).

Streamflow at each cross-section of a river is characterized by the discharge; i.e., the water volume passing through the cross-section in a unit time. Conventionally, the discharge has been considered a jump-driven stochastic process governed by a stochastic differential equation (SDE) of a mean-reversion type (Botter et al., 2013; Ceola et al., 2010). The SDE description enables us to analyze not only the streamflow dynamics but also the physical, biological, and chemical processes in river environments, such as the landscape control (Basso et al., 2015), the birth and death of riparian plants (Zen et al., 2020), and the water quality (Yoshioka and Yoshioka, 2021).

Discharge timeseries often exhibit a subexponential autocorrelation or a long memory due to the coexistence of fast and slow, namely multi-scale runoff processes, which cannot be captured by widely used linear autoregressive models (Ledvinka and Recknagel, 2020; Mihailović et al., 2019; Pelacani and Schmitt, 2021; Spezia et al., 2021). It has recently been found that the fundamental statistics (average, standard deviation, skewness, kurtosis, and histogram) and the subexponential autocorrelation function (ACF) of the discharge can be reasonably reproduced by a superposition of Ornstein–Uhlenbeck processes (a supOU process) (Yoshioka, 2022). The core of the supOU process, particularly the subexponential (more specifically, decaying in a power speed) ACF, is from the Lévy mixing mechanism, where processes having different time scales are efficiently superposed to obtain one multiscale model (Barndorff-Nielsen et al., 2013). The superposition is performed with respect to the mean reversion speed as the inverse of a characteristic time scale, yielding a non-Markovian SDE. Similar models have been applied to timeseries data in economics (Barndorff-Nielsen and Stelzer, 2013) and are well-studied in the extreme value theory (Moser and Stelzer, 2013). Directly handling a non-Markovian SDE is technically difficult in general; however, the Markovian embedding as a matrix analytic method potentially alleviates this difficulty. The Markovian embedding allows for reinterpreting a non-Markovian SDE as a system of (possibly infinitely many) Markovian SDEs and has been employed as an analytical tool for understanding and efficiently computing non-Markovian SDEs (Kanazawa et al., 2020; Loos et al., 2019; Abi Jaber, 2019; Mandjes and Sollie, 2021; Yoshioka and Tsujimura, 2022).

A common limitation of the abovementioned non-Markovian SDE is that it is not self-exciting as background driving noise processes are given exogenically. Nevertheless, some researchers have postulated that incorporating self-exciting effects or equivalently clustering effects works well in applied problems, including those arising in economics and the environment (Callegaro et al., 2022; Yoshioka and Tsujimura, 2022). Theoretically, this implies that one should use a state-dependent background driving jump process as in the continuous-state branching processes with immigration (the CBI process) (Li, 2020). Such an approach has been studied in the context of Volterra-type processes with state-dependent jump measures (Cuchiero and Teichmann, 2020), but not so for the SDEs of the abovementioned superposed type. Assessing the performance of such a superposed model in applications would be of interest.

*Objective and contribution*

The objectives of this work are the formulation, analysis, and engineering application of a superposition of affine jump processes of the continuous-state branching type, called a supCBI process, modeling the discharge timeseries of a river. This process is a generalization of the supOU process where Ornstein–Uhlenbeck processes are superposed instead of affine jump processes. The branching nature comes from the jump intensity affinely depending on the state variable, effectively representing the self-exciting jumps, which implies floods induced by clustered rainfall events in the context of discharge (Harka et al., 2021; Hussain et al., 2022; Yoshioka and Tsujimura, 2021). Focusing on an application, jumps in the supCBI process are assumed to be pure-jump subordinators.

The supCBI process is constructed heuristically based on a finite-dimensional system of mutually independent CBI processes of different reversion speeds. This system is integrated into the space of the reversion speed to obtain the supCBI process. Thus, the construction procedure of the supCBI process is based on a Markovian embedding. Convergence of the Markovian embedding in the sense of law is discussed through characteristic functions. The affine property of the supCBI process allows us to analytically obtain key statistical moments and the ACF. With the actual hourly discharge timeseries data, these statistics are used to identify the supCBI process and that the self-exciting effect better explains observed statistics than the supOU process.

We are also interested in model reduction as another application of the Markovian embedding, where a continuous-time and finite-state stochastic process is extracted from the supCBI process. The reduced model can be effectively used as a part of some integrated model having numerous state variables. In this study, we consider a process to dynamically classify flow regimes of discharge timeseries into two mutually exclusive regimes: a high-flow regime with a discharge higher than a prescribed threshold discharge and the opposite one called a low-flow regime. Such a binary classification problem arises in many applications, including but not limited to the sediment transport such that sand particles are transported along a watercourse when the discharge is higher than a threshold value (Downs and Soar, 2021), the stability analysis of riparian vegetation communities such that their extinction is ruled by a threshold discharge (Calvani et al., 2022), and the disaster management based on a prescribed threshold discharge above which severe flood damages may occur (Bischiniotis et al., 2020).

The high- and low-flow regimes are identified from the actual data using their waiting time distributions, from which we find that a phase-type distribution, such as the mixtures of exponential distributions, works well in explaining the waiting time statistics (Bladt and Nielsen, 2017). Phase-type distributions have been applied to problems related to hydrology (Bouzas et al., 2021; Fisher et al., 2010; Rho and Kim, 2019). However, to the best of our knowledge, their application to model reduction is rare.

For an in-depth analysis of supCBI process statistics numerically, an Euler–Maruyama-type Monte Carlo method (MCM) is proposed to simulate the supCBI process. This numerical method is based on a random number generator for tempered stable subordinators (Kawai and Masuda, 2011) and the discretization scheme for stationary affine processes of a superposed type (Stelzer et al., 2015). The computational results suggest the reasonable performance of the proposed MCM.

Consequently, we contribute to the modeling and application of a stochastic process with the help of Markovian embeddings. Our methodology is rather heuristic, but demonstrative computational results imply its applicability to engineering problems.

**Mathematical model**

*CBI processes*

We work with a complete probability space equipped with a sample space, event space, and a probability function as usual. We introduce a CBI process following the literature (Li, 2020; Yoshioka and Tsujimura, 2022). The time is denoted as $t \in \mathbb{R}$. In this study, a CBI process $Y = (Y_t)_{t \geq 0}$ is a unique solution to the SDE (e.g., Theorem 2.3 of Jin et al. (2020)):

$$dY_t = -\rho Y_t dt + \int_0^{A+\rho BY_{t-}} \int_0^\infty z\mu(du,dz,dt), \quad t > 0, \tag{1}$$

subject to an initial condition $Y_0 \geq 0$, where $Y_{t-}$ means the left-limit of $Y$ at $t$, $\rho > 0$ denotes the reversion speed, $A, B \geq 0$ are constant parameters that do not simultaneously become 0, and $\mu$ denotes a space–time Poisson random measure on $(0,+\infty) \times (0,+\infty) \times \mathbb{R}$ having the compensator $du \times v(dz) \times dt$ with the Lévy measure $v$ such that $\int_0^{+\infty} \min\{1,z\} v(dz) < +\infty$. We focus on the tempered stable case $v(dz) = e^{-bz} z^{-(\alpha+1)} dz$ with $b > 0$ and $\alpha < 1$, which is assumed in what follows for modeling simplicity and its applicability to the real data. Then, we obtain

$$M_k = \int_0^{+\infty} z^k v(dz) = b^{\alpha-k}\Gamma(k-\alpha) > 0, \quad k \in \mathbb{N}, \tag{2}$$

with $\Gamma(\cdot)$ as the Gamma function. The first and second terms of (1) physically represent the recession and occurrence of flood events, respectively. The formal jump intensity is proportional to the state-dependent coefficient, $A + \rho BY_{t-}$, representing the exogenic and self-exciting parts, respectively. The CBI process reduces to a classical jump-driven Ornstein–Uhlenbeck process when $B = 0$.

We assume the following to ensure the smallness of the self-exciting part

$$D \equiv 1 - BM_1 = 1 - Bb^{\alpha-1}\Gamma(1-\alpha) > 0, \tag{3}$$

which is crucial for the CBI process stationarity. This type of the boundedness of the self-exciting part is common in the affine jump processes like the hawkes process (Swishchuk et al., 2021). Under (3), the process $Y$ following (1) approaches to a stationary state by Theorem 2.7 of Jin et al. (2020). Indeed, consider the process $\left(\tilde{Y}_t\right)_{t\geq 0} = \left(A + \rho B Y_t\right)_{t\geq 0}$ to obtain the transformed SDE

$$\mathrm{d}\tilde{Y}_t = \rho\left(A - \tilde{Y}_t\right)\mathrm{d}t + \int_0^{\tilde{Y}_{t-}} \int_0^{\infty} z\hat{\mu}(\mathrm{d}u,\mathrm{d}z,\mathrm{d}t), \quad t > 0, \quad (4)$$

with which the admissibility condition is required for the theorem is satisfied with a Poisson random measure $\hat{\mu}(\mathrm{d}u,\mathrm{d}z,\mathrm{d}t) = \mu(\mathrm{d}u,\mathrm{d}z/\rho B,\mathrm{d}t)$.

At the stationary state, we have the stationary statistics (e.g., Yoshioka and Tsujimura, 2022), (For their derivations, see **Appendix A**):

$$\left(\text{Expectation of } Y\right) = \mathbb{E}[Y_t] = \frac{AM_1}{\rho D}, \quad (5)$$

$$\left(\text{Variance of } Y\right) = \mathbb{E}\left[\left(Y_t - \mathbb{E}[Y_t]\right)^2\right] = \frac{AM_2}{2\rho D^2}, \quad (6)$$

$$\left(\text{Skewness of } Y\right) = \frac{\mathbb{E}\left[\left(Y_t - \mathbb{E}[Y_t]\right)^3\right]}{\mathbb{E}\left[\left(Y_t - \mathbb{E}[Y_t]\right)^2\right]^{1.5}} = \frac{1}{\mathbb{E}\left[\left(Y_t - \mathbb{E}[Y_t]\right)^2\right]^{1.5}} \times A\left(\frac{M_3}{3\rho D^2} + \frac{1}{2}\frac{B(M_2)^2}{\lambda D^3}\right), \quad (7)$$

$$\left(\text{Kurtosis of } Y\right) = \frac{\mathbb{E}\left[\left(Y_t - \mathbb{E}[Y_t]\right)^4\right]}{\mathbb{E}\left[\left(Y_t - \mathbb{E}[Y_t]\right)^2\right]^2} - 3$$
$$= \frac{1}{\mathbb{E}\left[\left(Y_t - \mathbb{E}[Y_t]\right)^2\right]^2} \times \frac{A}{4\rho D}\left(\frac{M_4}{D} + 4\frac{BM_2M_3}{D^2} + 3\frac{B^2(M_2)^3}{D^3}\right). \quad (8)$$

In addition, the ACF $\mathrm{Cor}(s)$ of $Y$ for time lag $s \geq 0$ is given by

$$\mathrm{Cor}(s) = \exp(-\rho D s), \quad (9)$$

which decays exponentially at the rate $\rho D > 0$.

*Modeling supCBI process*

*Finite-dimensional model by Markovian embedding*

We heuristically construct a supCBI process starting from finite-dimensional Markovian SDEs. The CBI process with the reversion speed $\rho > 0$ is indicated by the superscript $(\rho)$. Let $n \in \mathbb{N}$, $\{\rho_i\}_{1\leq i \leq n}$ be a positive and strictly increasing sequence, and $\{c_i\}_{1\leq i \leq n}$ be a positive sequence with the partition-of-unity property $\sum_{i=1}^n c_i = 1$. Introduce the $n$-dimensional supCBI process $\{Y^{(\rho_i)}\}_{1\leq i \leq n}$ as a càdlàg solution to the system

$$\mathrm{d}Y_t^{(\rho_i)} = -\rho_i Y_t^{(\rho_i)} \mathrm{d}t + \int_0^{c_i A + \rho_i B Y_{t-}^{(\rho_i)}} \int_0^{\infty} z\mu_i(\mathrm{d}u,\mathrm{d}z,\mathrm{d}t), \quad t \in \mathbb{R}, \quad 1 \leq i \leq n. \quad (10)$$

Each $\mu_i$ ($1 \leq i \leq n$) is a mutually independent Poisson random measure extended to $t \in \mathbb{R}$ (Barndorff-Nielsen and Stelzer, 2013) having the compensator $\mathrm{d}u \times v(\mathrm{d}z) \times \mathrm{d}t$. With (10), the discharge $X_n = (X_{n,t})_{t \geq 0}$ is then defined as the discrete superposition

$$X_{n,t} = \underline{X} + \sum_{i=1}^{n} Y_t^{(\rho_i)}$$
$$= \underline{X} + \sum_{i=1}^{n} \int_0^{c_i A + \rho B Y_{t-}^{(\rho_i)}} \int_0^{+\infty} \int_{-\infty}^{t} \exp(-\rho_i(t-s)) z \mu_i(\mathrm{d}u, \mathrm{d}z, \mathrm{d}s), \quad t \in \mathbb{R}, \quad (11)$$

with a prescribed minimum discharge $\underline{X} > 0$.

Each $Y^{(\rho_i)}$ is independent with each other; hence, the system (10) admits a unique càdlàg strong solution. The contribution of each $Y^{(\rho_i)}$ is of the order $c_i$, as implied in the coefficient $c_i A$ and the moments of the CBI process presented in **Subsection 2.1**. Statistics of this system can be analytically obtained as presented in the next subsection.

For later use, let the probability measure be

$$\pi_n(\mathrm{d}\rho) = \sum_{i=1}^{n} c_i \delta_{\rho_i}, \quad \rho > 0, \quad (12)$$

where $\delta_{\rho_i}$ represents the Dirac's Delta concentrated at $\rho = \rho_i > 0$.

*supCBI process*

The finite-dimensional model in the previous subsection motivates us to consider an infinite-dimensional one as an integration of the continuum of the distributed process $\left[ Y^{(\rho)} \right]_{\rho > 0}$ for $\rho > 0$. We define a continuous counterpart $\pi(\mathrm{d}\rho)$ of the discrete probability density $\pi_n(\mathrm{d}\rho)$ for $\rho > 0$ such that

$$\int_0^{+\infty} \pi(\mathrm{d}\rho) = 1 \text{ and } \int_0^{+\infty} \frac{\pi(\mathrm{d}\rho)}{\rho} < +\infty. \quad (13)$$

The first condition is trivial, whereas the second equality is technical, analogous to that assumed in supOU processes (Stelzer et al., 2015), which enforces a regularity of $\pi$ near $\rho = 0$. We assume that $\pi$ has a smooth density function and a bounded average. In particular, we assume the Gamma density

$$\frac{\rho^{\beta-1}}{\Gamma(\beta)\eta^{\beta}} \exp\left(-\frac{\rho}{\eta}\right), \quad \rho > 0, \quad (14)$$

with the scaling parameter $\eta > 0$ and shape parameter $\beta > 1$. This density satisfies (13) and well fits to the real data in our application having polynomially decaying ACFs.

Now, we formally define the supCBI process. Invoking (11), we can formulate the supCBI process governing the discharge as the formal integral

$$X_t = \underline{X} + \int_0^{+\infty} Y_{t-}^{(\rho)}(\mathrm{d}\rho), \quad t \in \mathbb{R}. \quad (15)$$

with $\left( Y_t^{(\rho)}(\mathrm{d}\rho) \right)_{t \in \mathbb{R}}$ ($\rho > 0$) being measure-valued and is governed by the SDE

$$\mathrm{d}Y_t^{(\rho)}(\mathrm{d}\rho) = -\rho_i Y_t^{(\rho)}(\mathrm{d}\rho)\mathrm{d}t + \int_0^{A\pi(\mathrm{d}\rho)+\rho_i B Y_{t-}^{(\rho)}(\mathrm{d}\rho)} \int_0^\infty z \mu_\rho(\mathrm{d}u,\mathrm{d}z,\mathrm{d}t). \qquad (16)$$

Each $\mu_\rho$ ($\rho > 0$) is a mutually independent Poisson random measure for different $\rho$'s having the compensator $\mathrm{d}u \times v(\mathrm{d}z) \times \mathrm{d}t$. Our supCBI process is an ambit process with a high-dimensional Poisson random measure (Nguyen and Veraart, 2018). The supCBI process is seen as an infinite-dimensional system of SDEs with respect to the reversion speed having the probability density $\pi$. However, the meaning of (16) is vague because of the presence of the coefficient $A\pi(\mathrm{d}\rho)$ in the jump term, although the formal computation as performed in **Appendix A** may apply, as the average and variance of $Y_{t-}^{(\rho)}(\mathrm{d}\rho)$ are proportional to $\pi(\mathrm{d}\rho)$. In the next section, we reconsider the supCBI process as a weak limit of the finite-dimensional supCBI process in terms of characteristic functions.

*Remark 1* There exist other choices for the density $\pi$ to reproduce subexponential ACFs (Barndorff-Nielsen and Leonenko, 2005). However, they are analogous to (14) because their ACFs decay polynomially.

*Remark 2* The formal integration of the form (15) has already been suggested in the context of supOU processes (Barndorff-Nielsen, 2001). However, we did not find such a representation for the CBI processes.

**Analysis**

*Characteristic functions*

The consistency of characteristic functions between the finite- and infinite-dimensional supCBI processes is analyzed. By the mutual independence and affine property of $Y^{(\rho_i)}$ ($1 \leq i \leq n$), at time $t$ in a stationary state, the characteristic function of $X_n$ is as follows (Duffie et al., 2003):

$$C_n(u,t) \equiv \mathbb{E}\left[\exp(iuX_{n,t})\right] = e^{iu\underline{X}} \exp\left(A\sum_{i=1}^n c_i \int_0^{+\infty} \int_0^{+\infty} \left(\exp(i\phi_{i,s}z)-1\right)v(\mathrm{d}z)\mathrm{d}s\right), \quad u,t \in \mathbb{R}, \qquad (17)$$

with each $\phi_{i,s}$ ($1 \leq i \leq n$, $s \geq 0$) being a complex variable as a smooth solution to the initial value problem of the ordinary differential equation (ODE)

$$\mathrm{i}\frac{\mathrm{d}}{\mathrm{d}s}\phi_{i,s} = -\mathrm{i}\rho_i \phi_{i,s} + \rho_i B \int_0^{+\infty} \left(\exp(i\phi_{i,s}z)-1\right)v(\mathrm{d}z), \quad s > 0, \quad \phi_{i,0} = u. \qquad (18)$$

Note the independence of $C_n(u,t)$ on $t$ due to assuming the stationarity. Hence, we will omit its second argument for simplicity.

Similarly, we formally obtain a candidate of the characteristic function of $X$ at a stationary state as follows ($\mathrm{i}^2 = -1$):

$$C(u) \equiv e^{iu\underline{X}} \exp\left(A\int_0^{+\infty} \int_0^{+\infty} \int_0^{+\infty} \left(\exp(i\phi_s^{(\rho)}z)-1\right)v(\mathrm{d}z)\mathrm{d}s\pi(\mathrm{d}\rho)\right), \quad u,t \in \mathbb{R}, \qquad (19)$$

with each $\phi_s^{(\rho)}$ ( $\rho > 0$, $s \geq 0$ ) being a complex variable as a smooth solution to the initial value problem of the parameter-distributed ODE

$$i\frac{d}{ds}\phi_s^{(\rho)} = -i\rho\phi_s^{(\rho)} + \rho B \int_0^{+\infty} \left(\exp\left(i\phi_s^{(\rho)}z\right) - 1\right)v(dz), \quad s > 0, \quad \phi_0^{(\rho)} = u. \tag{20}$$

We want to define $C(u)$ in (19) as the characteristic function $\mathbb{E}\left[\exp(iuX_t)\right]$ of the supCBI process as a weak limit, limit in the sense of law, of its finite-dimensional Markovian embedding. The convergence of (17) to (19) should be analyzed to find in what sense the Markovian embedding provides an approximation to the supCBI process.

We analyze the properties of solutions $\phi_s^{(\rho)}$ to (20). First, we scale the time as $\tau = \rho s$ with a transformed variable $\phi_\tau = \phi_s^{(\rho)}$ to obtain

$$i\frac{d}{d\tau}\phi_\tau = -i\phi_\tau + B\int_0^{+\infty}\left(\exp(i\phi_\tau z) - 1\right)v(dz), \quad \tau > 0, \quad \phi_0 = u, \tag{21}$$

showing that solutions $\phi_\tau$ to (21) do not depend on $\rho$. Substituting $\phi_\tau = f_\tau + g_\tau i$ with some smooth real functions $f_\tau, g_\tau$ ( $\tau \geq 0$ ) into (21) yields

$$\frac{d}{d\tau}\left(if_\tau - g_\tau\right) = -if_\tau + g_\tau + B\int_0^{+\infty}\left(\exp(-g_\tau z)\left(\cos(f_\tau z) + i\sin(f_\tau z)\right) - 1\right)v(dz). \tag{22}$$

Then, we obtain a system of two ODEs for $\tau > 0$

$$\frac{d}{d\tau}f_\tau = -f_\tau + B\int_0^{+\infty}\exp(-g_\tau z)\sin(f_\tau z)v(dz), \tag{23}$$

$$\frac{d}{d\tau}g_\tau = -g_\tau + B\int_0^{+\infty}\left(1 - \exp(-g_\tau z)\cos(f_\tau z)\right)v(dz) \tag{24}$$

subject to initial values $f_0 = u$ and $g_0 = 0$. These equations are generalized Riccati equations of the CBI process (1), an affine process, with $\rho = 0$. A similar scaling with $\tau = \rho_i s$ applies to (17).

Assume $n \geq 2$ and a partition of the semi-infinite domain $(0, +\infty)$:

$$0 = x_0 < x_1 < ... < x_{n-1} < x_n = +\infty \tag{25}$$

and set

$$c_i = \int_{x_{i-1}}^{x_i}\pi(d\rho), \quad \rho_i = (x_{i-1} + x_i)/2 \quad (1 \leq i \leq n-1), \quad \rho_n = x_{n-1}. \tag{26}$$

This is a natural finite difference-like partition of the space of reversion speed (Yoshioka et al., 2022). To proceed, assume $x_i = \bar{C}in^{-\gamma}$ ( $1 \leq i \leq n-1$ ) with some constants $\bar{C} > 0$ and $\gamma \in (0,1)$, and $\int_{x_{i-1}}^{x_i}\frac{1}{\rho}\pi(d\rho) \leq C_0(x_i - x_{i-1})^\delta$ with some $C_0 > 0$ and $\delta \in (0,1]$. For example, if $\beta > 2$, we can choose $\delta = 1$ as the density function (14) becomes uniformly bounded. If $1 < \beta \leq 2$, we can choose $\delta = \beta - 1 \in (0,1]$.

We then obtain the proposition below whose proof is presented in **Appendix A**. With this proposition, we unbdertstand the supCBI process formally presented in the previous subsection as a limit in the sense of law of an appropriate sequence of finite-dimensional ones.

*Proposition 1*

Assume the partitions (25)–(26) with $x_i = \bar{C}in^{-\gamma}$ ( $1 \leq i \leq n-1$ ), $\bar{C} > 0$, $\gamma \in (0,1)$. We have $C_n(\cdot) \to C(\cdot)$ as $n \to +\infty$ at each $u \in \mathbb{R}$.

A corollary obtained from **Proposition 1** is the following convergence result of $R = \int_0^{+\infty} \frac{\pi(\mathrm{d}\rho)}{\rho}$.

*Corollary 1*

Under the partitions (25)–(26), we have $R_n \to R$ as $n \to +\infty$.

**Remark 3** The scaling of the jump rate $A + \rho B Y^{(\rho)}$ in $\rho$ plays an essential role in the analysis here. In addition, the assumption plays a role in calculating the statistics.

**Remark 4** Without assuming the stationarity, we obtain the characteristic function of the finite-dimensional supCBI process as

$$C_n(u,t) \equiv \mathbb{E}\left[\exp(iuX_{n,t}) \big| Y_0^{(\rho_i)} = y_i, i = 1,2,3,...,n\right]$$
$$= e^{iu\underline{X}} \exp\left(A\sum_{i=1}^n c_i \int_0^t \int_0^{+\infty} (\exp(i\phi_{i,s}z) - 1) v(\mathrm{d}z)\mathrm{d}s\right) \exp\left(i\sum_{i=1}^n \phi_{i,t} y_i\right), \quad u \in \mathbb{R}, \ t > 0 \qquad (27)$$

with $\phi_{i,t}$ being the unique smooth solution to the ODE

$$i\frac{\mathrm{d}}{\mathrm{d}s}\phi_{i,s} = -i\rho_i\phi_{i,s} + \rho_i B \int_0^{+\infty} (\exp(i\phi_{i,s}z) - 1) v(\mathrm{d}z), \quad s > 0, \ \phi_{i,0} = u. \qquad (28)$$

The last exponential of (27) is dropped out at the stationary limit $t \to +\infty$ due to $\lim_{t \to +\infty} \phi_{i,t} = 0$, with which (27) reduces to (17). The convergence analysis in the sense of law without assuming the stationarity will be more complicated due to the last exponential of (27) although (17) suffices for the analysis at stationary state. The non-stationary form (27) will be of importance when studying non-stationary supCBI processes having time-dependent coefficients like yearly-varying ones. Such models are beyond the scope of this work and will be considered in the future.

*Statistics*

The fundamental statistics of the supCBI process are analytically obtained, which are useful in both theory and application. We apply them to the analysis of streamflow timeseries in the next section. The derivation process discussed in this subsection is rather formal and starts from the finite-dimensional supCBI process, upon which we obtain the statistics of the infinite-dimensional one by letting $n \to +\infty$.

For the finite-dimensional case, we have the following proposition with the proof in **Appendix A**.

*Proposition 2*

Let $R_n = \sum_{i=1}^{n} \frac{c_i}{\rho_i}$. At a stationary state, we have

$$\mathbb{E}[X_n] = \underline{X} + \frac{AM_1}{D} R_n, \qquad (29)$$

$$\mathbb{E}\left[(X_n - \mathbb{E}[X_n])^2\right] = \frac{AM_2}{2D^2} R_n, \qquad (30)$$

$$\mathbb{E}\left[(X_n - \mathbb{E}[X_n])^3\right] = A\left(\frac{M_3}{3D^2} + \frac{1}{2}\frac{B(M_2)^2}{D^3}\right) R_n, \qquad (31)$$

$$\mathbb{E}\left[(X_n - \mathbb{E}[X_n])^4\right] - 3(\mathrm{Var}[X_n])^2 = \frac{A}{4D}\left(\frac{M_4}{D} + 4\frac{BM_2 M_3}{D^2} + 3\frac{B^2 (M_2)^3}{D^3}\right) R_n, \qquad (32)$$

$$\mathrm{Cor}_n(s) = \frac{1}{R_n} \sum_{i=1}^{n} \frac{c_i}{\rho_i} \exp(-\rho_i D s), \quad s \geq 0. \qquad (33)$$

The next proposition is an immediate consequence of **Corollary 1** and **Proposition 2**, where we can obtain the statistics of the supCBI process as a limit of its finite-dimensional version.

*Proposition 3*

Let $R = \int_0^{+\infty} \frac{\pi(\mathrm{d}\rho)}{\rho}$. Under the assumption of **Proposition 1**, we have

$$\mathbb{E}[X_n] \to \underline{X} + \frac{AM_1}{D} R, \qquad (34)$$

$$\mathbb{E}\left[(X_n - \mathbb{E}[X_n])^2\right] \to \frac{AM_2}{2D^2} R, \qquad (35)$$

$$\mathbb{E}\left[(X_n - \mathbb{E}[X_n])^3\right] \to \left(\frac{AM_3}{3D^2} + \frac{1}{2}\frac{AB(M_2)^2}{D^3}\right) R, \qquad (36)$$

$$\mathbb{E}\left[(X_n - \mathbb{E}[X_n])^4\right] - 3(\mathrm{Var}[X_n])^2 \to \frac{1}{4D}\left(\frac{AM_4}{D} + 4\frac{ABM_2 M_3}{D^2} + 3\frac{AB^2 (M_2)^3}{D^3}\right) R, \qquad (37)$$

$$\mathrm{Cor}_n(s) \to \frac{1}{R} \int_0^{+\infty} \frac{1}{\rho} \exp(-\rho D s) \pi(\mathrm{d}\rho), \quad s \geq 0. \qquad (38)$$

*Remark 5* The right-hand sides of (34)–(38) are the statistics of the supCBI process, which can also be obtained directly, as demonstrated in **Appendix A**. The error between each statistic is controlled by the difference $|R - R_n|$ quantified above, implying that the statistics in **Proposition 2** converge to the corresponding ones of **Proposition 1** if the characteristic function of the finite-dimensional supCBI process converges to that of the infinite-dimensional one.

**Application**

*Actual data*

The dataset we use in this study is the hourly discharge timeseries in mountainous perennial river environments in Japan. The study site is the two upstream rivers, Tabusa River and Jouge River, of Haizuka Dam, Gono River, Japan. Analyzing the discharge timeseries of these rivers is significant as they pour to Haizuka Dam, which is a specified multipurpose dam mainly operated for flood mitigation, supplying water resources to nearby cities, and discharging sustainable environmental flows of the downstream river environment (MLIT, http://www.cgr.mlit.go.jp/miyoshi/haizuka/about/index.html). The public hourly discharge data of two stations, called Stations 1 and 2, are available online from January 2017 to December 2020 (MLIT, http://www1.river.go.jp/cgi-bin/SiteInfo.exe?ID=307051287713030 for Station 1 in Tabusa River and http://www1.river.go.jp/cgi-bin/SiteInfo.exe?ID=307051287713010 for Station 2 in Jouge River). The data for Station 1 have been used by Yoshioka (2022). **Table 1** summarizes the average (Ave), standard deviation (Std), skewness (Skew), and kurtosis (Kurt) of the hourly data at Stations 1 and 2.

*Model fitting*

The model parameters of the supCBI process are identified at Stations 1 and 2. The proposed supCBI model is fitted against the actual data at the two stations as follows. The fitting method is a modification of those used by Yoshioka (2022) and Yoshioka et al. (2022) who analyzed supOU processes. First, the empirical ACF is used to fit the parameters $U = (1-BM_1)\eta = D\eta$ and $\beta$ of the density $\pi$ given a prescribed $D \in (0,1]$. As such, we can find $\eta$ as $UD^{-1}$. Then, the remaining parameters $A, B, b, \alpha$ are fitted so that the following normalized metric $\text{Er}^2$ containing relative errors (REs) of the following four statistics is minimized:

$$\text{Er}^2 = \left(\frac{\text{Ave}_M - \text{Ave}_M}{\text{Ave}_D}\right)^2 + \left(\frac{\text{Std}_M - \text{Std}_D}{\text{Std}_D}\right)^2 + \left(\frac{\text{Skew}_M - \text{Skew}_D}{\text{Skew}_D}\right)^2 + \left(\frac{\text{Kurt}_M - \text{Kurt}_D}{\text{Kurt}_D}\right)^2. \quad (39)$$

where the subscripts "M" and "D" mean the statistics obtained from the model and data, respectively. The reason for fixing $U$ is that numerical instability during the least-squares procedure when $U$ is smaller than around 0.5. We analyze the dependence of $U$ on estimation and computation later.

We consider an MCM for stationary supCBI processes. This MCM not only allows for computing statistics of supCBI processes but also the method itself is of interest because computational methods for non-Markovian processes have not been frequently discussed. The employed MCM is based on Yoshioka (2022) who numerically simulated sample paths of supOU processes at a stationary state. We extend this method to the supCBI process based on an ergodic assumption that each jump associates a reversion speed generated from the density $\pi$ (e.g., Stelzer et al., 2015). Based on these previous studies, the equation to temporally update $Y_t^{(\rho)}$ is heuristically set as

$$Y_{(n+1)\Delta t} = e^{-\rho_n \Delta t} Y_{n\Delta t} + \frac{1-e^{-\rho_n \Delta t}}{\rho_n \Delta t} \Delta L_{Y,n}, \quad n=0,1,2,... \quad (40)$$

with an initial condition $Y_0 > 0$, and that to compute the discharge $X$ is

$$X_{n\Delta t} = Y_{n\Delta t} + \underline{X}, \quad n=0,1,2,.... \quad (41)$$

The time increment $\Delta t$ (h) is chosen sufficiently small, each $\rho_n$ is i.i.d. and is sampled from the density proportional to $\frac{\pi}{\rho}$ considering the appearance of $R$ in the statistics of **Proposition 3** and $\Delta L_{Y,n}$ is a tempered stable process with the time increment $\Delta t$ having the local Lévy measure $(A + \rho_n Y_{n\Delta t} B) e^{-bz} z^{-(\alpha+1)} \mathrm{d}z$, $z > 0$. This MCM is heuristic, even for conventional CBI processes having a deterministic $\rho$. The rigorous convergence analysis of the numerical method is not within the scope of this paper, but we show that it reasonably works. Note that directly using the Markovian embedding maybe prohibitive as its computational cost is proportional to $n$.

**Table 2** summarizes the fitted parameter values at the two stations, where the computed statistics based on the MCM with sufficiently long timeseries ($\Delta t = 0.0005$ h, sampling duration 20,000 years) are also presented to check its validity. Statistics based on the identified model are presented in **Table 1**. **Figure 1** shows the empirical and fitted ACFs, where only the positive part of the empirical ACF was used to fit the model. The best least-squares results of the exponential ACF ($\exp(-\gamma s)$ with a parameter $\gamma > 0$) are also provided. The proposed model fits better than the exponential ACF. **Figure 2** compares the empirical and computed probability density functions (PDFs) of the discharge at each station. The obtained results suggest that the ACFs and statistics are well-fitted by the supCBI process. Further, the MCM successfully reproduces the statistics and PDF with reasonable accuracy. **Table 3** is provided to analyze the performance of the MCM, demonstrating its convergence, although it is irregular, attributable to the existence of the statistical bias.

A paramount finding from the fitting results is that the two stations have qualitatively different ACFs. Indeed, the ACF at Station 1 has a long memory ($\beta \in (1,2)$) where the integral of ACF with respect to the time lag diverges (e.g., Barndorff-Nielsen and Stelzer, 2013), whereas that of Station 2 does not $\beta > 2$. Nevertheless, both stations have $\beta$ close to 2. Although the data availability may affect the obtained results, they suggest that the nonexponential ACF is advantageous for modeling the discharge timeseries.

For a better understanding of the model performance, we analyze the dependence of the identified model parameters and the fitting error on the quantity $D = 1 - BM_1$, which was fixed before fitting the model. **Table 4** summarizes the identified parameter values and RE against different values of $D$ for Stations 2. The error metric $\mathrm{Er}^2$ of the fitting becomes smaller as $D$ increases, suggesting that it plays a key role in the model fitting procedure. All supCBI processes ($D < 1$) outperform the supOU process ($D = 1$) in terms of REs of the statistics, implying that the supCBI processes potentially serve as more accurate alternatives to the supOU process without significantly degrading the tractability. Specifying a smaller $D$ yields a larger $B$, allowing for the parameterization with a different degree of self-exciting nature.

*Model reduction*

We apply a model reduction to the identified supCBI processes, which is motivated by the fact that it will be useful if some simpler, as well as easier-to-use model, is extracted from a supCBI process. Indeed, the supCBI process is based on an infinite-dimensional SDE and may not be always suited to applied problems, such as optimal controls and differential games.

We extract semi-Markov chains as reduced models of the supCBI process. Exploring semi-Markovian models is a natural option because they potentially have a smaller degree of freedom than the supCBI process. We focus on the problem of finding reasonable two-regime semi-Markov chains representing the transition events from a high-flow regime to a low-flow regime, and vice versa. The two-regime description serves as the simplest model for dynamically classifying flow regimes where the threshold discharge governs the occurrence of physical phenomena, such as the sediment transport (Downs and Soar, 2021), biologically critical discharge above which the riparian vegetation communities extinct (Calvani et al., 2022), or a planning threshold of floods (Bischiniotis et al., 2020). In the study area, the physical or biological threshold would correspond to $O(10^0) - O(10^1)$ m³/s and the flood planning threshold to $O(10^1) - O(10^2)$ m³/s.

The high- and low-flow regimes are classified using a fixed threshold discharge $X_{\text{Thr}} > 0$, above and below which the flow is classified to be at the high- and low-flow regimes, respectively. The sequences of the stopping times $\{\theta_i\}_{i=1,2,3,...}$ and $\{\tau_i\}_{i=1,2,3,...}$ to classify the two regimes are set as follows (**Figure 3**):

$$\theta_i = \inf\{t > 0 | X_t > X_{\text{Thr}}, \theta_i > \theta_{i-1}\}, \quad \theta_0 = -\infty, \quad i = 1, 2, 3, ..., \tag{42}$$

$$\tau_i = \inf\{t > 0 | X_t < X_{\text{Thr}}, \tau_i > \tau_{i-1}\}, \quad \tau_0 = -\infty, \quad i = 1, 2, 3, .... \tag{43}$$

The values of $\theta_0, \tau_0$ are chosen for brevity, where we set the initial time as 0. Based on the two sequences, the sequences of the waiting times during high- and low-flow regimes are determined as $\{\tau_i - \theta_i\}_{i=1,2,3,...}$ and $\{\theta_{i+1} - \tau_i\}_{i=1,2,3,...}$, respectively. The waiting times can be computed by the MCM once $X_{\text{Thr}}$ is fixed. The computational resolution and duration are the same as those employed in the previous subsection. For fitting the probability distributions, we again use (39) but omit the last term on the kurtosis, which turns out to work reasonably well. This is because the number of parameters will be smaller than that in the previous subsection, and we encountered a numerical instability when containing the original error metric.

We focus on the model reduction by the mixture of exponential distributions as the simplest phase-type model for generating semi-Markovian waiting times (Bladt and Nielsen, 2017):

$$p(Z) = w_1 \lambda_1 e^{-\lambda_1 Z} + w_2 \lambda_2 e^{-\lambda_2 Z}, \quad Z \geq 0, \tag{44}$$

with parameters $\lambda_1 \geq \lambda_2 > 0$ and $w_1 + w_2 = 1$ such that $w_1, w_2 \geq 0$, which reduces to the exponential distribution if $\lambda_1 = \lambda_2$. One may use a mixture of a large number of exponential distributions whose approximability of completely monotone functions has been verified (Merkle, 2014). We focus on the minimal case (44) containing at most two exponentials. There is no a priori theoretical verification on the uniqueness and existence of the minimizer of (39). In some cases, the proposed fitting method reasonably works in several cases, as discussed below.

**Table 5** shows the least-squares estimations of $\lambda_1, \lambda_2, w_1, w_2$ for different values of the threshold discharge $X_{\text{Thr}}$ for Station 1. **Table 6** shows the corresponding fitting results. Similarly, **Tables 7** and **8** summarize the fitted results for Station 2. **Figures 4** and **5** compare the PDFs of waiting times in the two regimes for Stations 1 and 2, respectively. The fitting results for the low-flow regime with relatively small $X_{\text{Thr}}$ show that the statistical moments can be accurately fitted by a mixture of two exponential distributions, whereas the use of single exponential distribution performs better for the low-flow regime.

The PDFs computed by the supCBI process seem to be continuous, decreasing, and convex, supporting their complete monotonicity and hence the use of the mixture of exponential distributions. Indeed, the PDF computed using the supCBI process is captured by the reduced model for the high-flow regime. Nevertheless, a clear discrepancy is observed in the small waiting time at which the PDFs by the supCBI process have a sharp decrease. The reduced model does not necessarily capture this variation.

Finally, we consider the model reduction for different values of $D$ focusing on a particular case of $X_{\text{Thr}} = 20$ m³/s at Station 2. **Table 9** shows the least-squares estimations of the four parameters $w_1, w_2, \lambda_1, \lambda_2$ against different values of $D$. According to **Table 9**, we infer that the error metric of the reduced model becomes smaller as the value of $D$ decreases, namely, as the degree of self-exciting nature increases, implying that the mixture of exponential distributions explains the waiting times of the high- and low-flow regimes as the underlying supCBI process becomes more strongly self-exciting. We consider that introducing the two possibly different time scales, $\lambda_1^{-1}, \lambda_2^{-1}$, results in the reasonable performance of the proposed semi-Markovian model reduction. Finally, we note that increasing the number of exponential functions to be mixed would improves the performance of the model reduction, but increases the degrees of freedom and degrades its efficiency.

**Conclusions**

In this study, we presented supCBI process as a jump-driven stochastic process. Their formulation was based on a Markovian embedding, with which the major statistics of interest were obtained analytically. The supCBI processes were fitted to discharge data collected at two rivers, demonstrating that they serve as an effective stochastic process model of the streamflow dynamics. Semi-Markov chains describing transitions between the binary flow regimes were successfully obtained from the supCBI processes.

We proposed a mathematical framework for superposing jump-driven affine processes, which would apply to jump-diffusion processes as well. Such a problem may arise when considering the coupled dynamics of discharge and water quality indices and aquatic communities (Yoshioka et al., 2021; Wang et al., 2021). The semi-Markov chains as reduced models of the supCBI processes can be efficiently used as parsimonious tools for river flows in integrated environmental management (Cho et al., 2021; Schmitz et al., 2016) possibly including hydropower generation (Jiang et al., 2022).

The research on supCBI processes is still at an initial stage, and there are several unresolved issues, including a sharper analysis of the convergence of the finite-dimensional model, measure change, and model ambiguity (Wu et al., 2022), stochastic control (Lü and Zhang, 2021), and extreme value statistics (Soulier, 2021), all of which are related to management problems of aquatic environments and ecosystems. The path-wise convergence of the Markovian embedding, which was not addressed in this study, should also be analyzed in the future. The convergence of an MCM used in this study is also unresolved. An advanced MCM, if it will become available, can then be utilized for extracting a reduced model using the proposed semi-Markovian approach. Currently, we are tackling these issues by focusing on the management of an existing dam-downstream river environment.

The superposition considered in this work is just one example. There would be many other alternative, such that the jump term depends on the superposed process itself. This problem will be addressed through applications to real data. Analysis without assuming the stationarity will also be of interest.

**Acknowledgements**
A part of computation of this paper was carried out using the supercomputer of ACCMS, Kyoto University, Japan.

**Funding**
Japan Society for the Promotion of Science (22K14441), Kurita Water and Environment Foundation (21K008), and Environmental Research Projects from the Sumitomo Foundation (203160) support this research.

**Competing interests**
The author declares that there are no competing interests in this paper.

**Table 1.** Empirical, fitted, and computed statistics at each station. For the supCBI process, we set $D = 0.7$.

|  |  | Ave (m$^3$/s) | Std (m$^3$/s) | Skew (-) | Kurt (-) |
|---|---|---|---|---|---|
| Station 1 | Empirical | 2.593.E+00 | 7.835.E+00 | 1.609.E+01 | 4.049.E+02 |
|  | Model | 2.578.E+00 | 7.878.E+00 | 1.487.E+01 | 4.176.E+02 |
| Station 2 | Empirical | 2.504.E+00 | 7.258.E+00 | 1.053.E+01 | 1.617.E+02 |
|  | Model | 2.485.E+00 | 7.310.E+00 | 9.790.E+00 | 1.663.E+02 |

**Table 2.** Fitted parameter values at each station ($D = 0.7$).

| Parameter | Station 1 | Station 2 |
|---|---|---|
| $U$ (1/h) | 1.08.E-01 | 6.76.E-02 |
| $\beta$ (-) | 1.75.E+00 | 2.04.E+00 |
| $\alpha$ (-) | 7.20.E-01 | 4.56.E-01 |
| $b$ (s/m$^3$) | 8.31.E-03 | 1.76.E-02 |
| $A$ (m$^{3\alpha}$/s$^{\alpha}$/h) | 1.70.E-02 | 1.16.E-02 |
| $B$ (m$^{3(\alpha-1)}$/s$^{(\alpha-1)}$/h) | 2.44.E-02 | 2.04.E-02 |
| $\underline{X}$ (m$^3$/s) | 0.00.E-02 | 6.00.E-02 |

**Table 3.** Convergence of the statistics by the MCM at Station 2.

| Statistics |  | Ave (m$^3$/s) | Std (m$^3$/s) | Skew (-) | Kurt (-) |
|---|---|---|---|---|---|
| Theory | Value | 2.485.E+00 | 7.310.E+00 | 9.790.E+00 | 1.663.E+02 |
| $\Delta t = 0.0020$ (h) | Value | 2.476.E+00 | 7.280.E+00 | 9.696.E+00 | 1.622.E+02 |
| 5,000 years | Relative error | 3.672.E-03 | 4.186.E-03 | 9.771.E-03 | 2.538.E-02 |
| $\Delta t = 0.0067$ (h) | Value | 2.489.E+00 | 7.322.E+00 | 9.728.E+00 | 1.657.E+02 |
| 15,000 years | Relative error | 1.401.E-03 | 1.684.E-03 | 6.460.E-03 | 3.537.E-03 |
| $\Delta t = 0.0005$ (h) | Value | 2.486.E+00 | 7.303.E+00 | 9.738.E+00 | 1.663.E+02 |
| 20,000 years | Relative error | 2.640.E-04 | 9.095.E-04 | 5.388.E-03 | 3.481.E-05 |
| $\Delta t = 0.00033$ (h) | Value | 2.487.E+00 | 7.313.E+00 | 9.717.E+00 | 1.640.E+02 |
| 30,000 years | Relative error | 6.701.E-04 | 3.676.E-04 | 7.561.E-03 | 1.391.E-02 |

**Table 4.** The identified parameter values and the RE against different values of $D$ for Station 2.

| $D$ (-) | 1.0 | 0.9 | 0.8 | 0.7 | 0.6 | 0.5 |
|---|---|---|---|---|---|---|
| $\alpha$ (-) | 2.86.E-01 | 3.40.E-01 | 3.96.E-01 | 4.56.E-01 | 5.19.E-01 | 5.87.E-01 |
| $b$ (s/m$^3$) | 1.61.E-02 | 1.66.E-02 | 1.71.E-02 | 1.76.E-02 | 1.82.E-02 | 1.88.E-02 |
| $A$ (m$^{3\alpha}$/s$^\alpha$/h) | 6.98.E-03 | 8.30.E-03 | 9.82.E-03 | 1.16.E-02 | 1.34.E-02 | 1.53.E-02 |
| $B$ (m$^{3(\alpha-1)}$/s$^{(\alpha-1)}$/h) | 0.00.E+00 | 4.89.E-03 | 1.16.E-02 | 2.04.E-02 | 3.16.E-02 | 4.51.E-02 |
| RE of Ave | 8.97.E-03 | 8.41.E-03 | 7.91.E-03 | 7.42.E-03 | 6.95.E-03 | 6.49.E-03 |
| RE of St | 8.67.E-03 | 8.09.E-03 | 7.61.E-03 | 7.14.E-03 | 6.69.E-03 | 6.25.E-03 |
| RE of Skew | 8.08.E-02 | 7.73.E-02 | 7.37.E-02 | 7.02.E-02 | 6.65.E-02 | 6.28.E-02 |
| RE of Kurt | 3.18.E-02 | 3.07.E-02 | 2.95.E-02 | 2.82.E-02 | 2.69.E-02 | 2.56.E-02 |
| Er$^2$ | 7.69.E-03 | 7.05.E-03 | 6.43.E-03 | 5.82.E-03 | 5.24.E-03 | 4.68.E-03 |

**Table 5.** Least-squares estimations of the four parameters $w_1, w_2, \lambda_1, \lambda_2$ against different values of the threshold discharge $X_{Thr} > 0$ for Station 1.

| $X_{Thr}$ (m$^3$/s) | Regime | $w_1$ (-) | $w_2$ (-) | $\lambda_1$ (1/h) | $\lambda_2$ (1/h) |
|---|---|---|---|---|---|
| 5 | High | 8.167.E-01 | 1.833.E-01 | 9.391.E-02 | 1.040.E+06 |
|   | Low | 3.271.E-01 | 6.729.E-01 | 2.792.E-01 | 8.163.E-03 |
| 20 | High | 4.672.E-01 | 5.328.E-01 | 1.416.E-01 | 1.416.E-01 |
|    | Low | 2.739.E-01 | 7.261.E-01 | 4.514.E-01 | 2.075.E-03 |
| 50 | High | 5.631.E-02 | 9.437.E-01 | 3.645.E+06 | 1.828.E-01 |
|    | Low | 2.708.E-01 | 7.292.E-01 | 1.155.E+03 | 6.710.E-04 |
| 100 | High | 6.447.E-02 | 9.355.E-01 | 2.231.E+05 | 2.405.E-01 |
|     | Low | 2.781.E-01 | 7.219.E-01 | 3.860.E-01 | 2.126.E-04 |

**Table 6.** The statistics by the supCBI process and mixture of exponential distributions (MixExp) against different values of the threshold discharge $X_{\text{Thr}} > 0$ for Station 1.

| $X_{\text{Thr}}$ (m³/s) | Regime | Class | Ave (h) | Std (h) | Skew (-) | Kurt (-) | Er² |
|---|---|---|---|---|---|---|---|
| 5 | High | supCBI | 8.573.E+00 | 1.063.E+01 | 2.028.E+00 | 5.493.E+00 | 1.425.E-03 |
| | | MixExp | 8.697.E+00 | 1.047.E+01 | 2.092.E+00 | 6.417.E+00 | |
| | Low | supCBI | 8.360.E+01 | 1.150.E+02 | 2.319.E+00 | 7.769.E+00 | 5.233.E-10 |
| | | MixExp | 8.360.E+01 | 1.150.E+02 | 2.319.E+00 | 7.626.E+00 | |
| 20 | High | supCBI | 6.688.E+00 | 7.531.E+00 | 1.766.E+00 | 3.922.E+00 | 2.462.E-02 |
| | | MixExp | 7.060.E+00 | 7.060.E+00 | 2.000.E+00 | 6.000.E+00 | |
| | Low | supCBI | 3.505.E+02 | 4.630.E+02 | 2.207.E+00 | 7.004.E+00 | 2.386.E-12 |
| | | MixExp | 3.505.E+02 | 4.630.E+02 | 2.207.E+00 | 6.999.E+00 | |
| 50 | High | supCBI | 5.023.E+00 | 5.626.E+00 | 1.740.E+00 | 3.727.E+00 | 2.553.E-02 |
| | | MixExp | 5.162.E+00 | 5.461.E+00 | 2.009.E+00 | 6.038.E+00 | |
| | Low | supCBI | 1.085.E+03 | 1.437.E+03 | 2.191.E+00 | 6.717.E+00 | 1.425.E-05 |
| | | MixExp | 1.087.E+03 | 1.435.E+03 | 2.198.E+00 | 6.950.E+00 | |
| 100 | High | supCBI | 3.787.E+00 | 4.273.E+00 | 1.770.E+00 | 3.923.E+00 | 2.025.E-02 |
| | | MixExp | 3.890.E+00 | 4.150.E+00 | 2.012.E+00 | 6.050.E+00 | |
| | Low | supCBI | 3.397.E+03 | 4.519.E+03 | 2.209.E+00 | 6.960.E+00 | 1.685.E-09 |
| | | MixExp | 3.397.E+03 | 4.519.E+03 | 2.209.E+00 | 7.009.E+00 | |

**Table 7.** Least-squares estimations of the four parameters $w_1, w_2, \lambda_1, \lambda_2$ against different values of the threshold discharge $X_{\text{Thr}} > 0$ for Station 2.

| $X_{\text{Thr}}$ (m³/s) | Regime | $w_1$ | $w_2$ | $\lambda_1$ (1/h) | $\lambda_2$ (1/h) |
|---|---|---|---|---|---|
| 5 | High | 5.336.E-01 | 4.664.E-01 | 8.547.E-02 | 8.547.E-02 |
| | Low | 2.513.E-01 | 7.487.E-01 | 1.891.E-01 | 7.887.E-03 |
| 20 | High | 5.018.E-01 | 4.982.E-01 | 1.351.E-01 | 1.351.E-01 |
| | Low | 2.743.E-01 | 7.257.E-01 | 2.371.E-01 | 2.335.E-03 |
| 50 | High | 4.950.E-01 | 5.050.E-01 | 2.033.E-01 | 2.035.E-01 |
| | Low | 3.093.E-01 | 6.907.E-01 | 2.689.E+01 | 6.809.E-04 |
| 100 | High | 1.616.E-01 | 8.384.E-01 | 9.759.E+00 | 2.710.E-01 |
| | Low | 3.043.E-01 | 6.957.E-01 | 2.689.E+01 | 1.633.E-04 |

**Table 8.** The statistics by the supCBI process and MixExp against different values of the threshold discharge $X_{\text{Thr}} > 0$ for Station 2.

| $X_{\text{Thr}}$ (m³/s) | Regime | Class | Ave (h) | Std (h) | Skew (-) | Kurt (-) | $Er^2$ (-) |
|---|---|---|---|---|---|---|---|
| 5 | High | supCBI | 1.125.E+01 | 1.223.E+01 | 1.857.E+00 | 4.786.E+00 | 9.452.E-03 |
| | | MixExp | 1.170.E+01 | 1.170.E+01 | 2.000.E+00 | 6.000.E+00 | |
| | Low | supCBI | 9.626.E+01 | 1.217.E+02 | 2.209.E+00 | 7.104.E+00 | 2.620.E-12 |
| | | MixExp | 9.626.E+01 | 1.217.E+02 | 2.209.E+00 | 7.015.E+00 | |
| 20 | High | supCBI | 7.122.E+00 | 7.733.E+00 | 1.870.E+00 | 4.841.E+00 | 8.234.E-03 |
| | | MixExp | 7.402.E+00 | 7.402.E+00 | 2.000.E+00 | 6.000.E+00 | |
| | Low | supCBI | 3.119.E+02 | 4.109.E+02 | 2.213.E+00 | 6.980.E+00 | 2.213.E-12 |
| | | MixExp | 3.119.E+02 | 4.109.E+02 | 2.213.E+00 | 7.031.E+00 | |
| 50 | High | supCBI | 4.665.E+00 | 5.233.E+00 | 1.956.E+00 | 5.191.E+00 | 7.088.E-03 |
| | | MixExp | 4.917.E+00 | 4.917.E+00 | 2.000.E+00 | 6.000.E+00 | |
| | Low | supCBI | 1.021.E+03 | 1.388.E+03 | 2.278.E+00 | 7.691.E+00 | 1.648.E-04 |
| | | MixExp | 1.014.E+03 | 1.397.E+03 | 2.257.E+00 | 7.270.E+00 | |
| 100 | High | supCBI | 3.110.E+00 | 3.628.E+00 | 2.091.E+00 | 6.004.E+00 | 5.248.E-10 |
| | | MixExp | 3.110.E+00 | 3.628.E+00 | 2.091.E+00 | 6.415.E+00 | |
| | Low | supCBI | 4.253.E+03 | 5.842.E+03 | 2.243.E+00 | 7.101.E+00 | 1.026.E-05 |
| | | MixExp | 4.260.E+03 | 5.833.E+03 | 2.249.E+00 | 7.224.E+00 | |

**Table 9.** The identified parameter values of the reduced model with the corresponding error metric; we consider Station 2 with $X_{\text{Thr}} = 20$ m³/s.

| $D$ | Regime | Class | Ave (h) | Std (h) | Skew (-) | Kurt (-) | $\text{Er}^2$ (-) |
|---|---|---|---|---|---|---|---|
| 1.0 | High | supCBI | 9.886.E+00 | 8.192.E+00 | 1.084.E+00 | 1.050.E+00 | 7.317.E-01 |
| | | MixExp | 8.882.E+00 | 8.882.E+00 | 2.000.E+00 | 6.000.E+00 | |
| | Low | supCBI | 4.262.E+02 | 4.559.E+02 | 2.013.E+00 | 6.010.E+00 | 5.587.E-06 |
| | | MixExp | 4.264.E+02 | 4.557.E+02 | 2.017.E+00 | 6.073.E+00 | |
| 0.9 | High | supCBI | 8.970.E+00 | 8.098.E+00 | 1.340.E+00 | 2.137.E+00 | 2.480.E-01 |
| | | MixExp | 8.489.E+00 | 8.489.E+00 | 2.000.E+00 | 6.000.E+00 | |
| | Low | supCBI | 3.884.E+02 | 4.438.E+02 | 2.063.E+00 | 6.247.E+00 | 3.455.E-12 |
| | | MixExp | 3.884.E+02 | 4.438.E+02 | 2.063.E+00 | 6.277.E+00 | |
| 0.8 | High | supCBI | 8.050.E+00 | 7.926.E+00 | 1.595.E+00 | 3.362.E+00 | 6.462.E-02 |
| | | MixExp | 7.987.E+00 | 7.987.E+00 | 2.000.E+00 | 6.000.E+00 | |
| | Low | supCBI | 3.499.E+02 | 4.277.E+02 | 2.128.E+00 | 6.607.E+00 | 2.762.E-12 |
| | | MixExp | 3.499.E+02 | 4.277.E+02 | 2.128.E+00 | 6.596.E+00 | |
| 0.7 | High | supCBI | 7.122.E+00 | 7.733.E+00 | 1.870.E+00 | 4.841.E+00 | 8.234.E-03 |
| | | MixExp | 7.402.E+00 | 7.402.E+00 | 2.000.E+00 | 6.000.E+00 | |
| | Low | supCBI | 3.119.E+02 | 4.109.E+02 | 2.213.E+00 | 6.980.E+00 | 2.213.E-12 |
| | | MixExp | 3.119.E+02 | 4.109.E+02 | 2.213.E+00 | 7.031.E+00 | |
| 0.6 | High | supCBI | 6.207.E+00 | 7.446.E+00 | 2.148.E+00 | 6.508.E+00 | 3.189.E-12 |
| | | MixExp | 6.207.E+00 | 7.446.E+00 | 2.148.E+00 | 6.698.E+00 | |
| | Low | supCBI | 2.731.E+02 | 3.907.E+02 | 2.332.E+00 | 7.614.E+00 | 2.321.E-12 |
| | | MixExp | 2.731.E+02 | 3.907.E+02 | 2.332.E+00 | 7.696.E+00 | |
| 0.5 | High | supCBI | 5.270.E+00 | 7.056.E+00 | 2.472.E+00 | 8.719.E+00 | 2.271.E-12 |
| | | MixExp | 5.270.E+00 | 7.056.E+00 | 2.472.E+00 | 8.617.E+00 | |
| | Low | supCBI | 2.335.E+02 | 3.672.E+02 | 2.509.E+00 | 8.715.E+00 | 2.196.E-12 |
| | | MixExp | 2.335.E+02 | 3.672.E+02 | 2.509.E+00 | 8.767.E+00 | |

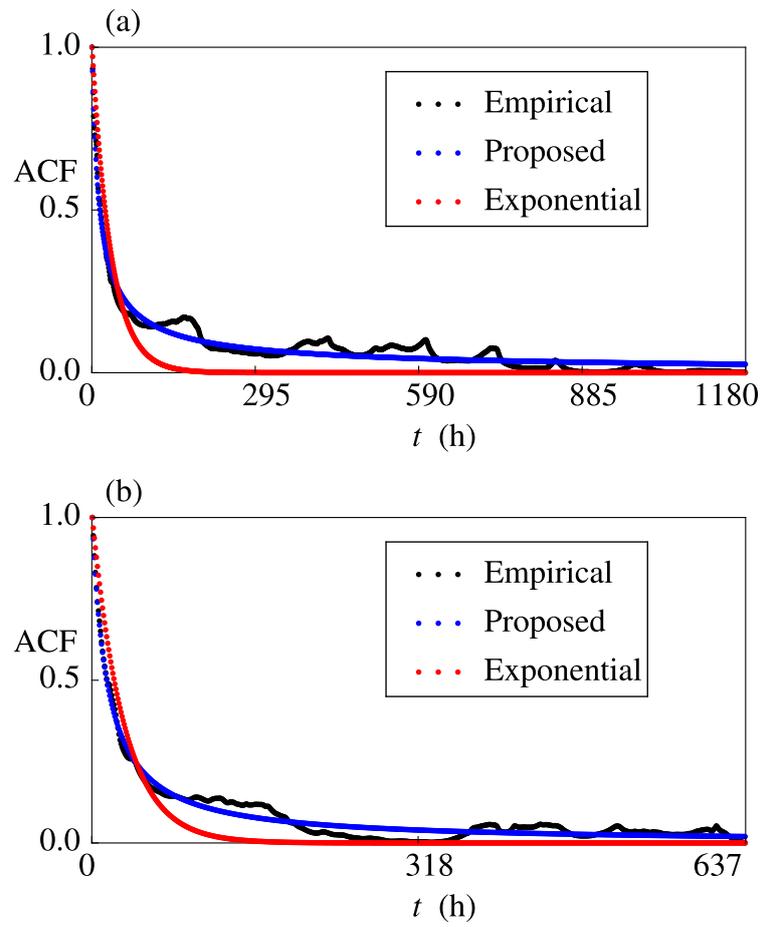

**Figure 1.** Empirical and fitted ACFs at (a) Station 1 and (b) Station 2.

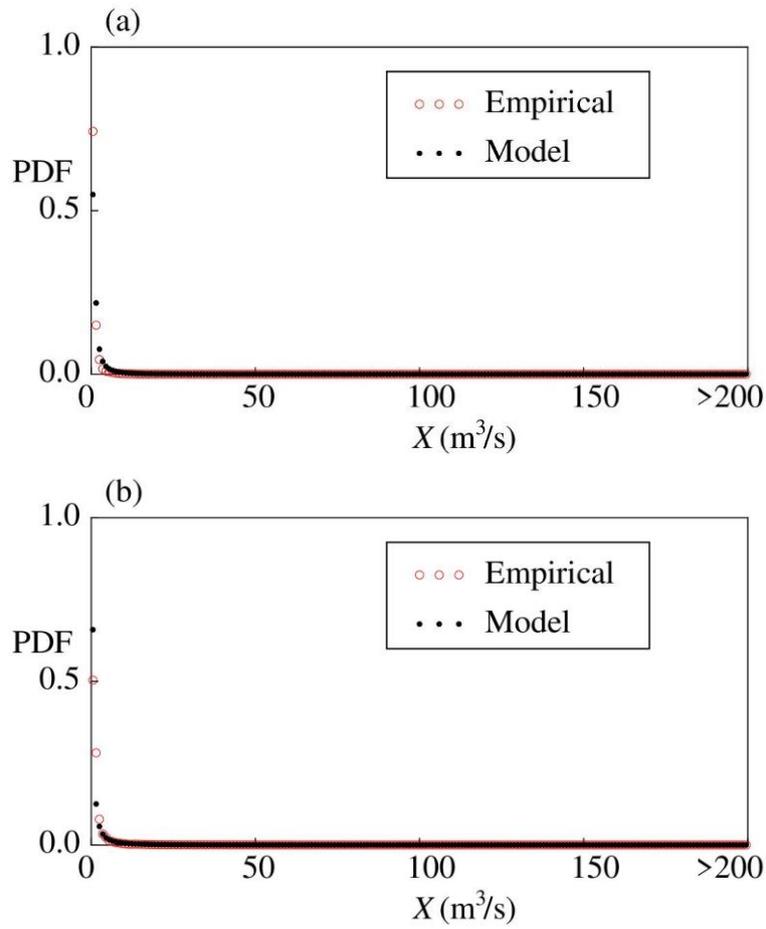

**Figure 2.** Empirical and computed PDFs at (a) Station 1 and (b) Station 2.

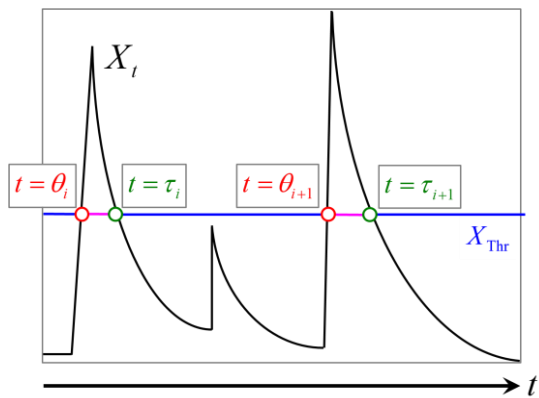

**Figure 3.** A conceptual figure of the model reduction. With the two sequences of the crossing times $\{\theta_i\}_{i=1,2,3,...}$ and $\{\tau_i\}_{i=1,2,3,...}$.

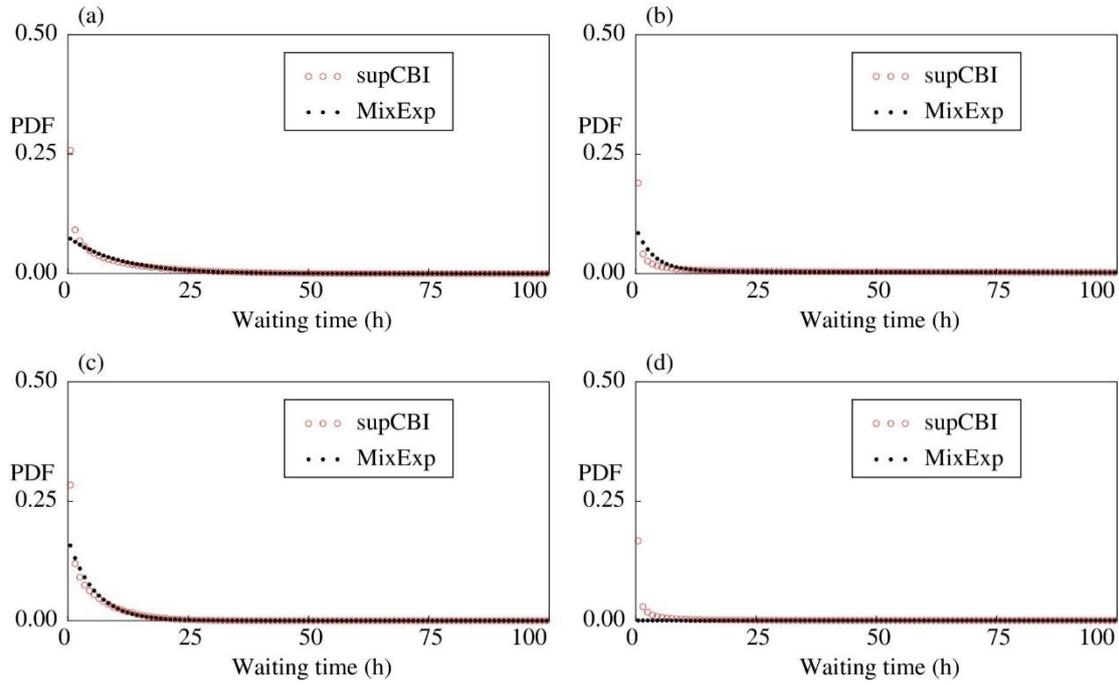

**Figure 4.** Comparison of the PDFs of waiting times for high- and low-flow regimes against different values of the threshold discharge $X_{\text{Thr}}$ (m$^3$/s) for Station 1: (a and b) high- and low-flow regimes, respectively, with $X_{\text{Thr}} = 5$ m$^3$/s and (c and d) high- and low-flow regimes, respectively, with $X_{\text{Thr}} = 50$ m$^3$/s.

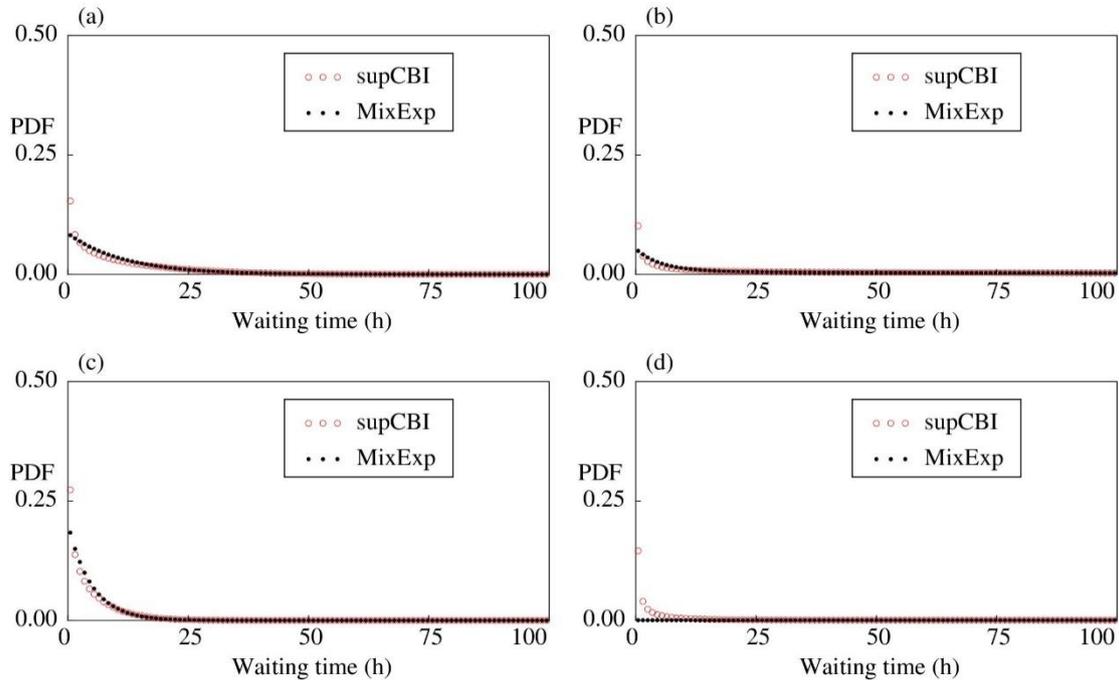

**Figure 5.** Comparison of the PDFs of waiting times for high- and low-flow regimes against different values of the threshold discharge $X_{\text{Thr}}$ for Station 2: (a and b) high- and low-flow regimes, respectively, with $X_{\text{Thr}} = 5$ m$^3$/s and (c and d) high- and low-flow regimes, respectively, with $X_{\text{Thr}} = 50$ m$^3$/s.

**Appendix A: calculation procedures of statistics**

We present calculation procedures of the statistics of the CBI and supCBI processes. As a supCBI process is a superposition of independent CBI processes, the central part of the calculation procedure depends essentially on the statistical properties of the latter. The proofs of some propositions are presented in this appendix as well.

*A.1 Statistics of CBI process*

We express the CBI process in a stationary state as follows:

$$dY_t = -\rho Y_t dt + dN_Y \tag{A.1}$$

with a space–time Poisson random measure $N_Y$ with the formal jump intensity $(A + B\rho Y_{t-})v(dz)$. The subscript representing time is hereafter omitted when there is no confusion. For $k \in \mathbb{N}$, let $m_k = \mathbb{E}[Y^k]$. The methodology used is adapted from a reprint of Yoshioka and Tsujimura (2022) based on Bernis and Scotti (2020) with a significant addition of the detailed calculation procedure.

First, by taking the expectation, we calculate the average $m_1$ based on (A.1) as follows (This amounts to evaluating the steady state of the differential equation of the corresponding moment):

$$0 = -\rho m_1 + (A + B\rho m_1)M_1, \tag{A.2}$$

namely,

$$m_1 = \frac{AM_1}{\rho(1-BM_1)} = \frac{AM_1}{\rho D}. \tag{A.3}$$

We obtain $m_2, m_3, m_4$ similarly. On the variance, by the definition, we have

$$V = \mathbb{E}\left[(Y-m_1)^2\right] = m_2 - (m_1)^2. \tag{A.4}$$

From (A.1), we obtain

$$d(Y^2) = -2\rho Y^2 dt + z^2 + 2Yz. \tag{A.5}$$

where $z$ denotes the jump size at time $t$. With (A.3), we proceed as follows:

$$0 = -2\rho m_2 + (A + B\rho m_1)M_2 + 2(Am_1 + B\rho m_2)M_1, \tag{A.6}$$

$$2\rho(1-BM_1)\left(V + \left(\frac{AM_1}{\rho(1-BM_1)}\right)^2\right) = \frac{AM_2}{1-BM_1} + 2\frac{(AM_1)^2}{\rho(1-BM_1)}, \tag{A.7}$$

thus, we have

$$V = \frac{AM_2}{2\rho(1-BM_1)^2} = \frac{AM_2}{2\rho D^2}. \tag{A.8}$$

On the skewness, we set

$$\begin{aligned}
\overline{S} &= \mathbb{E}\left[(Y-m_1)^3\right] \\
&= m_3 + 2(m_1)^3 - 3m_1 m_2 \\
&= m_3 + 2(m_1)^3 - 3m_1\left(V + (m_1)^2\right) \\
&= m_3 - (m_1)^3 - 3m_1 V
\end{aligned} \tag{A.9}$$

and obtain

$$m_3 = \overline{S} + (m_1)^3 + 3m_1 V. \tag{A.10}$$

From the SDE (A.1), we have

$$\mathrm{d}(Y^3) = -3\rho Y^3 \mathrm{d}t + z^3 + 3Yz^2 + 3Y^2 z. \tag{A.11}$$

Hence, by taking the expectation, we have

$$0 = -3\rho m_3 + (A + B\rho m_1)M_3 + 3(Am_1 + B\rho m_2)M_2 + 3(Am_2 + B\rho m_3)M_1 \tag{A.12}$$

thus, we obtain

$$\begin{aligned}
&3\rho(1-BM_1)\left(\overline{S} + (m_1)^3 + 3m_1 V\right) \\
&= \frac{AM_3}{1-BM_1} + 3\left(Am_1 + B\rho\left(V + (m_1)^2\right)\right)M_2 + 3AM_1\left(V + (m_1)^2\right).
\end{aligned} \tag{A.13}$$

We proceed as follows:

$$\begin{aligned}
&3\left(Am_1 + B\rho\left(V + (m_1)^2\right)\right)M_2 \\
&= 3M_2\left(A\frac{AM_1}{\rho(1-BM_1)} + B\rho\left(\frac{AM_2}{2\rho(1-BM_1)^2} + \left(\frac{AM_1}{\rho(1-BM_1)}\right)^2\right)\right) \\
&= 3\frac{A^2 M_1 M_2}{\rho(1-BM_1)} + \frac{3}{2}\frac{AB(M_2)^2}{(1-BM_1)^2} + 3\frac{B(AM_1)^2 M_2}{\rho(1-BM_1)^2} \\
&= 3\frac{A^2 M_1 M_2}{\rho(1-BM_1)^2} - 3\frac{A^2 B(M_1)^2 M_2}{\rho(1-BM_1)^2} + \frac{3}{2}\frac{AB(M_2)^2}{(1-BM_1)^2} + 3\frac{A^2 B(M_1)^2 M_2}{\rho(1-BM_1)^2} \\
&= 3\frac{A^2 M_1 M_2}{\rho(1-BM_1)^2} + \frac{3}{2}\frac{AB(M_2)^2}{(1-BM_1)^2}
\end{aligned} \tag{A.14}$$

$$\begin{aligned}
3AM_1\left(V + (m_1)^2\right) &= 3AM_1\left(\frac{AM_2}{2\rho(1-BM_1)^2} + \left(\frac{AM_1}{\rho(1-BM_1)}\right)^2\right) \\
&= \frac{3}{2}\frac{A^2 M_1 M_2}{\rho(1-BM_1)^2} + 3\frac{(AM_1)^3}{\rho^2(1-BM_1)^2}
\end{aligned} \tag{A.15}$$

and

$$
\begin{aligned}
&-3\rho(1-BM_1)\left((m_1)^3+3m_1V\right) \\
&=-3\rho(1-BM_1)\left(\left(\frac{AM_1}{\rho(1-BM_1)}\right)^3+3\frac{AM_1}{\rho(1-BM_1)}\frac{AM_2}{2\rho(1-BM_1)^2}\right). \\
&=-3\frac{(AM_1)^3}{\rho^2(1-BM_1)^2}-\frac{9}{2}\frac{A^2M_1M_2}{\rho(1-BM_1)^2}
\end{aligned}
\tag{A.16}
$$

Consequently, we have (Some terms are colored as in the equation below; the colored terms can be canceled out or merged.)

$$
\begin{aligned}
3\rho(1-BM_1)\bar{S}&=\frac{AM_3}{1-BM_1} \\
&+3\Big(Am_1+B\rho\big(V+(m_1)^2\big)\Big)M_2+3AM_1\big(V+(m_1)^2\big) \\
&-3\rho(1-BM_1)\big((m_1)^3+3m_1V\big) \\
&=\frac{AM_3}{1-BM_1} \\
&+3\frac{A^2M_1M_2}{\rho(1-BM_1)^2}+\frac{3}{2}\frac{AB(M_2)^2}{(1-BM_1)^2}+\frac{3}{2}\frac{A^2M_1M_2}{\rho(1-BM_1)^2}+3\frac{(AM_1)^3}{\rho^2(1-BM_1)^2} \\
&-3\frac{(AM_1)^3}{\rho^2(1-BM_1)^2}-\frac{9}{2}\frac{A^2M_1M_2}{\rho(1-BM_1)^2} \\
&=\frac{AM_3}{1-BM_1}+\frac{3}{2}\frac{AB(M_2)^2}{(1-BM_1)^2}
\end{aligned}
\tag{A.17}
$$

namely,

$$
\bar{S}=\frac{1}{3\rho(1-BM_1)}\left(\frac{AM_3}{1-BM_1}+\frac{3}{2}\frac{AB(M_2)^2}{(1-BM_1)^2}\right)=\frac{AM_3}{3\rho D^2}+\frac{1}{2}\frac{AB(M_2)^2}{\rho D^3}.
\tag{A.18}
$$

On the kurtosis, we set

$$
\begin{aligned}
\bar{K}&=\mathbb{E}\left[(Y-m_1)^4\right] \\
&=m_4-4m_3m_1+6m_2(m_1)^2-3(m_1)^4 \\
&=m_4-4\big(\bar{S}+(m_1)^3+3m_1V\big)m_1+6\big(V+(m_1)^2\big)(m_1)^2-3(m_1)^4 \\
&=m_4-4\bar{S}m_1-4(m_1)^4-12V(m_1)^2+6V(m_1)^2+6(m_1)^4-3(m_1)^4 \\
&=m_4-4\bar{S}m_1-6V(m_1)^2-(m_1)^4
\end{aligned}
\tag{A.19}
$$

From the SDE (A.1), we obtain

$$
\mathrm{d}(Y^4)=-4\rho Y^4\mathrm{d}t+z^4+4Yz^3+6Y^2z^2+4Y^3z,
\tag{A.20}
$$

from which we obtain

$$
\begin{aligned}
0&=-4\rho m_4+(A+B\rho m_1)M_4+4(Am_1+B\rho m_2)M_3 \\
&+6(Am_2+B\rho m_3)M_2+4(Am_3+B\rho m_4)M_1
\end{aligned}
\tag{A.21}
$$

hence, we have

$$4\rho(1-BM_1)m_4 = (A+B\rho m_1)M_4 + 4(Am_1+B\rho m_2)M_3 \qquad (A.22)$$
$$+6(Am_2+B\rho m_3)M_2 + 4Am_3M_1$$

Thus, we have

$$4\rho(1-BM_1)\left(\bar{K}+4\bar{S}m_1+6V(m_1)^2+(m_1)^4\right) = (A+B\rho m_1)M_4$$
$$+4(Am_1+B\rho m_2)M_3 \qquad . \qquad (A.23)$$
$$+6(Am_2+B\rho m_3)M_2 + 4Am_3M_1$$

We rewrite (A.23) as follows:

$$4\rho(1-BM_1)\bar{K}$$
$$= (A+B\rho m_1)M_4 + 4\left(Am_1+B\rho\left(V+(m_1)^2\right)\right)M_3$$
$$+6\left(A\left(V+(m_1)^2\right)+B\rho\left(\bar{S}+(m_1)^3+3m_1V\right)\right)M_2 + 4A\left(\bar{S}+(m_1)^3+3m_1V\right)M_1 \qquad (A.24)$$
$$-4\rho(1-BM_1)\left(4\bar{S}m_1+6V(m_1)^2+(m_1)^4\right)$$

Each term of (A.24) is calculated as follows:

$$(A+B\rho m_1)M_4 = \frac{AM_4}{1-BM_1}, \qquad (A.25)$$

$$4\left(Am_1+B\rho\left(V+(m_1)^2\right)\right)M_3$$
$$= 4M_3\left(A\frac{AM_1}{\rho(1-BM_1)}+B\rho\left(\frac{AM_2}{2\rho(1-BM_1)^2}+\left(\frac{AM_1}{\rho(1-BM_1)}\right)^2\right)\right)$$
$$= 4\frac{A^2M_1M_3}{\rho(1-BM_1)} + 2\frac{ABM_2M_3}{(1-BM_1)^2} + 4\frac{A^2B(M_1)^2M_3}{\rho(1-BM_1)^2} \qquad , \qquad (A.26)$$
$$= 4\frac{A^2M_1M_3}{\rho(1-BM_1)^2} - 4\frac{A^2B(M_1)^2M_3}{\rho(1-BM_1)^2} + 2\frac{ABM_2M_3}{(1-BM_1)^2} + 4\frac{A^2B(M_1)^2M_3}{\rho(1-BM_1)^2}$$
$$= 4\frac{A^2M_1M_3}{\rho(1-BM_1)^2} + 2\frac{ABM_2M_3}{(1-BM_1)^2}$$

$$6A\left(V+(m_1)^2\right)M_2 = 6AM_2\left(\frac{AM_2}{2\rho(1-BM_1)^2}+\left(\frac{AM_1}{\rho(1-BM_1)}\right)^2\right),$$
$$= 3\frac{(AM_2)^2}{\rho(1-BM_1)^2} + 6\frac{A^3(M_1)^2M_2}{\rho^2(1-BM_1)^2} \qquad (A.27)$$

$$6B\rho\left(\bar{S}+\left(m_{1}\right)^{3}+3m_{1}V\right)M_{2}$$

$$=6B\rho M_{2}\left(\begin{array}{c}\dfrac{AM_{3}}{3\rho\left(1-BM_{1}\right)^{2}}+\dfrac{1}{2}\dfrac{AB\left(M_{2}\right)^{2}}{\rho\left(1-BM_{1}\right)^{3}}+\left(\dfrac{AM_{1}}{\rho\left(1-BM_{1}\right)}\right)^{3}\\ +3\dfrac{AM_{1}}{\rho\left(1-BM_{1}\right)}\dfrac{AM_{2}}{2\rho\left(1-BM_{1}\right)^{2}}\end{array}\right),\quad\text{(A.28)}$$

$$=2\dfrac{ABM_{2}M_{3}}{\left(1-BM_{1}\right)^{2}}+3\dfrac{AB^{2}\left(M_{2}\right)^{3}}{\left(1-BM_{1}\right)^{3}}+6\dfrac{A^{3}B\left(M_{1}\right)^{3}M_{2}}{\rho^{2}\left(1-BM_{1}\right)^{3}}+9\dfrac{A^{2}BM_{1}\left(M_{2}\right)^{2}}{\rho\left(1-BM_{1}\right)^{3}}$$

$$4A\left(\bar{S}+\left(m_{1}\right)^{3}+3m_{1}V\right)M_{1}$$

$$=4AM_{1}\left(\begin{array}{c}\dfrac{AM_{3}}{3\rho\left(1-BM_{1}\right)^{2}}+\dfrac{1}{2}\dfrac{AB\left(M_{2}\right)^{2}}{\rho\left(1-BM_{1}\right)^{3}}+\left(\dfrac{AM_{1}}{\rho\left(1-BM_{1}\right)}\right)^{3}\\ +3\dfrac{AM_{1}}{\rho\left(1-BM_{1}\right)}\dfrac{AM_{2}}{2\rho\left(1-BM_{1}\right)^{2}}\end{array}\right),\quad\text{(A.29)}$$

$$=\dfrac{4}{3}\dfrac{A^{2}M_{1}M_{3}}{\rho\left(1-BM_{1}\right)^{2}}+2\dfrac{A^{2}BM_{1}\left(M_{2}\right)^{2}}{\rho\left(1-BM_{1}\right)^{3}}+4\dfrac{A^{4}\left(M_{1}\right)^{4}}{\rho^{3}\left(1-BM_{1}\right)^{3}}+6\dfrac{A^{3}\left(M_{1}\right)^{2}M_{2}}{\rho^{2}\left(1-BM_{1}\right)^{3}}$$

and

$$4\rho\left(1-BM_{1}\right)\left(4\bar{S}m_{1}+6V\left(m_{1}\right)^{2}+\left(m_{1}\right)^{4}\right)$$

$$=4\rho\left(1-BM_{1}\right)\left(\begin{array}{c}4\left(\dfrac{AM_{3}}{3\rho\left(1-BM_{1}\right)^{2}}+\dfrac{1}{2}\dfrac{AB\left(M_{2}\right)^{2}}{\rho\left(1-BM_{1}\right)^{3}}\right)\dfrac{AM_{1}}{\rho\left(1-BM_{1}\right)}\\ +6\dfrac{AM_{2}}{2\rho\left(1-BM_{1}\right)^{2}}\left(\dfrac{AM_{1}}{\rho\left(1-BM_{1}\right)}\right)^{2}+\left(\dfrac{AM_{1}}{\rho\left(1-BM_{1}\right)}\right)^{4}\end{array}\right).\quad\text{(A.30)}$$

$$=4\rho\left(1-BM_{1}\right)\left(\begin{array}{c}\dfrac{4A^{2}M_{1}M_{3}}{3\rho^{2}\left(1-BM_{1}\right)^{3}}+2\dfrac{A^{2}BM_{1}\left(M_{2}\right)^{2}}{\rho^{2}\left(1-BM_{1}\right)^{4}}\\ +3\dfrac{A^{3}\left(M_{1}\right)^{2}M_{2}}{\rho^{3}\left(1-BM_{1}\right)^{4}}+\dfrac{A^{4}\left(M_{1}\right)^{4}}{\rho^{4}\left(1-BM_{1}\right)^{4}}\end{array}\right)$$

$$=\dfrac{16}{3}\dfrac{A^{2}M_{1}M_{3}}{\rho\left(1-BM_{1}\right)^{2}}+8\dfrac{A^{2}BM_{1}\left(M_{2}\right)^{2}}{\rho\left(1-BM_{1}\right)^{3}}+12\dfrac{A^{3}\left(M_{1}\right)^{2}M_{2}}{\rho^{2}\left(1-BM_{1}\right)^{3}}+4\dfrac{A^{4}\left(M_{1}\right)^{4}}{\rho^{3}\left(1-BM_{1}\right)^{3}}$$

Now, by (A.25)–(A.30), we obtain (again, some terms are colored in the equation below; the colored terms can be canceled out or merged.)

$$4\rho(1-BM_1)\bar{K}$$
$$=\frac{AM_4}{1-BM_1}+4\frac{A^2M_1M_3}{\rho(1-BM_1)^2}+2\frac{ABM_2M_3}{(1-BM_1)^2}$$
$$+3\frac{(AM_2)^2}{\rho(1-BM_1)^2}+6\frac{A^3(M_1)^2M_2}{\rho^2(1-BM_1)^2}$$
$$+2\frac{ABM_2M_3}{(1-BM_1)^2}+3\frac{AB^2(M_2)^3}{(1-BM_1)^3}+6\frac{A^3B(M_1)^3M_2}{\rho^2(1-BM_1)^3}+9\frac{A^2BM_1(M_2)^2}{\rho(1-BM_1)^3}$$
$$+\frac{4}{3}\frac{A^2M_1M_3}{\rho(1-BM_1)^2}+2\frac{A^2BM_1(M_2)^2}{\rho(1-BM_1)^3}+4\frac{A^4(M_1)^4}{\rho^3(1-BM_1)^3}+6\frac{A^3(M_1)^2M_2}{\rho^2(1-BM_1)^3}$$
$$-\frac{16}{3}\frac{A^2M_1M_3}{\rho(1-BM_1)^2}-8\frac{A^2BM_1(M_2)^2}{\rho(1-BM_1)^3}-12\frac{A^3(M_1)^2M_2}{\rho^2(1-BM_1)^3}-4\frac{A^4(M_1)^4}{\rho^3(1-BM_1)^3}$$
$$=\frac{AM_4}{1-BM_1}+4\frac{ABM_2M_3}{(1-BM_1)^2}+3\frac{(AM_2)^2}{\rho(1-BM_1)^2}+6\frac{A^3(M_1)^2M_2}{\rho^2(1-BM_1)^2}$$
$$+3\frac{AB^2(M_2)^3}{(1-BM_1)^3}+6\frac{A^3B(M_1)^3M_2}{\rho^2(1-BM_1)^3}+3\frac{A^2BM_1(M_2)^2}{\rho(1-BM_1)^3}-6\frac{A^3(M_1)^2M_2}{\rho^2(1-BM_1)^3}\,, \quad (A.31)$$

which leads to

$$4\rho(1-BM_1)\bar{K}$$
$$=\frac{AM_4}{1-BM_1}+4\frac{ABM_2M_3}{(1-BM_1)^2}+3\frac{(AM_2)^2}{\rho(1-BM_1)^3}-3\frac{A^2BM_1(M_2)^2}{\rho(1-BM_1)^3}$$
$$+6\frac{A^3(M_1)^2M_2}{\rho^2(1-BM_1)^3}-6\frac{A^3B(M_1)^3M_2}{\rho^2(1-BM_1)^3}+3\frac{AB^2(M_2)^3}{(1-BM_1)^3}+6\frac{A^3B(M_1)^3M_2}{\rho^2(1-BM_1)^3}, \quad (A.32)$$
$$+3\frac{A^2BM_1(M_2)^2}{\rho(1-BM_1)^3}-6\frac{A^3(M_1)^2M_2}{\rho^2(1-BM_1)^3}$$
$$=\frac{AM_4}{1-BM_1}+3\frac{(AM_2)^2}{\rho(1-BM_1)^3}+4\frac{ABM_2M_3}{(1-BM_1)^2}+3\frac{AB^2(M_2)^3}{(1-BM_1)^3}$$

This equality is rewritten as follows:

$$4\rho(1-BM_1)\bar{K}=\frac{AM_4}{1-BM_1}+3\frac{(AM_2)^2}{\rho(1-BM_1)^3}+4\frac{ABM_2M_3}{(1-BM_1)^2}+3\frac{AB^2(M_2)^3}{(1-BM_1)^3} \quad (A.33)$$

and we obtain

$$\bar{K}=\frac{1}{4\rho(1-BM_1)}\left(\frac{AM_4}{1-BM_1}+3\frac{(AM_2)^2}{\rho(1-BM_1)^3}+4\frac{ABM_2M_3}{(1-BM_1)^2}+3\frac{AB^2(M_2)^3}{(1-BM_1)^3}\right)$$
$$=3\left(\frac{AM_2}{2\rho(1-BM_1)^2}\right)^2+\frac{1}{4\rho(1-BM_1)}\left(\frac{AM_4}{1-BM_1}+4\frac{ABM_2M_3}{(1-BM_1)^2}+3\frac{AB^2(M_2)^3}{(1-BM_1)^3}\right). \quad (A.34)$$
$$=3\mathrm{Var}[Y]^2+\frac{A}{4\rho(1-BM_1)}\left(\frac{M_4}{1-BM_1}+4\frac{BM_2M_3}{(1-BM_1)^2}+3\frac{B^2(M_2)^3}{(1-BM_1)^3}\right)$$

Thus, we have

$$\bar{K} - 3\operatorname{Var}[Y]^2 = \frac{1}{4\rho D}\left(\frac{AM_4}{D} + 4\frac{ABM_2 M_3}{D^2} + 3\frac{AB^2(M_2)^3}{D^3}\right). \tag{A.35}$$

Finally, we obtain the autocorrelation with a time lag $s \geq 0$. At time $t$ under a stationary state, we formally have

$$\mathrm{d}(Y_t Y_{t+s}) = Y_t \mathrm{d}(Y_{t+s}) = -\rho Y_t Y_{t+s}\mathrm{d}s + Y_t z_{t+s}, \tag{A.36}$$

where the increment $\mathrm{d}$ is considered with respect to $s$. Then, we obtain

$$\begin{aligned}\frac{\mathrm{d}}{\mathrm{d}s}\mathbb{E}[Y_t Y_{t+s}] &= -\rho \mathbb{E}[Y_t Y_{t+s}]\mathrm{d}s + AM_1 \mathbb{E}[Y_t] + \rho BM_1 \mathbb{E}[Y_t Y_{t+s}]\\ &= -\rho(1 - BM_1)\mathbb{E}[Y_t Y_{t+s}] + AM_1 m_1\end{aligned}. \tag{A.37}$$

Therefore, by (A.3),

$$\begin{aligned}&\frac{\mathrm{d}}{\mathrm{d}s}\mathbb{E}[Y_t Y_{t+s}]\\ &= \frac{\mathrm{d}}{\mathrm{d}s}\mathbb{E}\left[Y_t Y_{t+s} - (m_1)^2\right]\\ &= -\rho(1 - BM_1)\mathbb{E}\left[Y_t Y_{t+s} - (m_1)^2\right] - \rho(1 - BM_1)(m_1)^2 + AM_1 m_1\\ &= -\rho(1 - BM_1)\mathbb{E}\left[Y_t Y_{t+s} - (m_1)^2\right]\end{aligned}. \tag{A.38}$$

Consequently, we obtain

$$\frac{\mathbb{E}\left[Y_t Y_{t+s} - (m_1)^2\right]}{\mathbb{E}\left[(Y_t)^2 - (m_1)^2\right]} = \exp(-\rho(1 - BM_1)s) = \exp(-\rho D s). \tag{A.39}$$

Similar technique has been applied to related self-exciting jump models (Proposition 7 of Eyjolfsson and Tjøstheim, 2021). Our result is consistent with it.

### *A.2 Statistics of finite-dimensional supCBI process (A proof of Proposition 2)*

We derive the statistics of the finite-dimensional supCBI process $X_n$ ($n \in \mathbb{N}$) based on the results obtained in **Subsection A.1**. A statistic of a CBI process with the reversion speed $\rho$ is indicated by the superscript $(\rho)$. For later use, let $R_n = \sum_{i=1}^{n}\frac{c_i}{\rho_i}$. Recall that the jump rate of $Y^{(\rho_i)}$ is proportional to $c_i A + B\rho_i Y^{(\rho_i)}$.

On the average, because a finite-dimensional supCBI process is a weighted sum of mutually independent CBI processes, we obtain

$$\mathbb{E}[X_n] = \underline{X} + \sum_{i=1}^{n} m_1^{(\rho_i)} = \underline{X} + \sum_{i=1}^{n}\frac{c_i AM_1}{\rho_i(1 - BM_1)} = \underline{X} + \frac{AM_1}{1 - BM_1}R_n = \underline{X} + \frac{AM_1}{D}R_n. \tag{A.40}$$

On the variance, we have

$$\begin{aligned}
\mathrm{Var}[X_n] &= \mathbb{E}\left[\left(X_n - \mathbb{E}[X_n]\right)^2\right] \\
&= \mathbb{E}\left[\left(\sum_{i=1}^{n}\left(Y^{(\rho_i)} - m_1^{(\rho_i)}\right)\right)^2\right] \\
&= \mathbb{E}\left[\sum_{i=1}^{n}\left(Y^{(\rho_i)} - m_1^{(\rho_i)}\right)^2\right] \\
&= \sum_{i=1}^{n}\mathbb{E}\left[\left(Y^{(\rho_i)} - m_1^{(\rho_i)}\right)^2\right]
\end{aligned} \tag{A.41}$$

because $Y^{(\rho_i)} - m_1^{(\rho_i)}$ has the mean 0 and each $Y^{(\rho_i)} - m_1^{(\rho_i)}$ and $Y^{(\rho_j)} - m_1^{(\rho_j)}$ are mutually independent when $i \neq j$. Hence, we obtain

$$\mathrm{Var}[X_n] = \sum_{i=1}^{n}\frac{c_i AM_2}{2\rho_i(1-BM_1)^2} = \frac{AM_2}{2(1-BM_1)^2}R_n = \frac{AM_2}{2D^2}R_n. \tag{A.42}$$

On the skewness, we have

$$\mathbb{E}\left[\left(X_n - \mathbb{E}[X_n]\right)^3\right] = \mathbb{E}\left[\left(\sum_{i=1}^{n}\left(Y^{(\rho_i)} - m_1^{(\rho_i)}\right)\right)^3\right] = \sum_{i=1}^{n}\mathbb{E}\left[\left(Y^{(\rho_i)} - m_1^{(\rho_i)}\right)^3\right] \tag{A.43}$$

thus, we obtain

$$\begin{aligned}
\mathbb{E}\left[\left(X_n - \mathbb{E}[X_n]\right)^3\right] &= \sum_{i=1}^{n}\left(\frac{c_i AM_3}{3\rho_i(1-BM_1)^2} + \frac{1}{2}\frac{c_i AB(M_2)^2}{\rho_i(1-BM_1)^3}\right) \\
&= \left(\frac{AM_3}{3D^2} + \frac{1}{2}\frac{AB(M_2)^2}{D^3}\right)R_n
\end{aligned} \tag{A.44}$$

On the kurtosis, again by the independence

$$\begin{aligned}
&\mathbb{E}\left[\left(X_n - \mathbb{E}[X_n]\right)^4\right] \\
&= \mathbb{E}\left[\left(\sum_{i=1}^{n}\left(Y^{(\rho_i)} - m_1^{(\rho_i)}\right)\right)^4\right] \\
&= \sum_{i=1}^{n}\mathbb{E}\left[\left(Y^{(\rho_i)} - m_1^{(\rho_i)}\right)^4\right] + 6\mathbb{E}\left[\sum_{i \neq j}^{n}\left(Y^{(\rho_i)} - m_1^{(\rho_i)}\right)^2\left(Y^{(\rho_j)} - m_1^{(\rho_j)}\right)^2\right] \\
&= \sum_{i=1}^{n}\mathbb{E}\left[\left(Y^{(\rho_i)} - m_1^{(\rho_i)}\right)^4\right] + 6\sum_{i \neq j}^{n}\mathbb{E}\left[\left(Y^{(\rho_i)} - m_1^{(\rho_i)}\right)^2\right]\mathbb{E}\left[\left(Y^{(\rho_j)} - m_1^{(\rho_j)}\right)^2\right] \\
&= \sum_{i=1}^{n}\mathbb{E}\left[\left(Y^{(\rho_i)} - m_1^{(\rho_i)}\right)^4\right] - 3\sum_{i=1}^{n}\left(\mathbb{E}\left[\left(Y^{(\rho_i)} - m_1^{(\rho_i)}\right)^2\right]\right)^2 + 3\left(\sum_{i=1}^{n}\mathbb{E}\left[\left(Y^{(\rho_i)} - m_1^{(\rho_i)}\right)^2\right]\right)^2 \\
&= \sum_{i=1}^{n}\mathbb{E}\left[\left(Y^{(\rho_i)} - m_1^{(\rho_i)}\right)^4\right] - 3\sum_{i=1}^{n}\left(\mathbb{E}\left[\left(Y^{(\rho_i)} - m_1^{(\rho_i)}\right)^2\right]\right)^2 + 3\left(\mathbb{E}\left[\sum_{i=1}^{n}\left(Y^{(\rho_i)} - m_1^{(\rho_i)}\right)^2\right]\right)^2 \\
&= \sum_{i=1}^{n}\left\{\mathbb{E}\left[\left(Y^{(\rho_i)} - m_1^{(\rho_i)}\right)^4\right] - \mathrm{Var}\left[Y^{(\rho_i)}\right]\right\} + 3\left(\mathrm{Var}[X_n]\right)^2
\end{aligned} \tag{A.45}$$

we obtain

$$\begin{aligned}
&\mathbb{E}\left[\left(X_n - \mathbb{E}[X_n]\right)^4\right] - 3\left(\operatorname{Var}[X_n]\right)^2 \\
&= \sum_{i=1}^{n}\left\{\mathbb{E}\left[\left(Y^{(\rho_i)} - m_1^{(\rho_i)}\right)^4\right] - 3\operatorname{Var}\left[Y^{(\rho_i)}\right]\right\} \\
&= \sum_{i=1}^{n} \frac{c_i}{4\rho_i(1-BM_1)}\left(\frac{AM_4}{1-BM_1} + 4\frac{ABM_2 M_3}{(1-BM_1)^2} + 3\frac{AB^2(M_2)^3}{(1-BM_1)^3}\right) \\
&= \frac{1}{4D}\left(\frac{AM_4}{D} + 4\frac{ABM_2 M_3}{D^2} + 3\frac{AB^2(M_2)^3}{D^3}\right) R_n
\end{aligned} \quad (A.46)$$

Finally, on the autocorrelation, at a stationary state, we have

$$\begin{aligned}
\mathbb{E}\left[X_{n,t}X_{n,t+s} - \left(\mathbb{E}[X_{n,t}]\right)^2\right] &= \mathbb{E}\left[\left(X_{n,t} - \mathbb{E}[X_{n,t}]\right)\left(X_{n,t+s} - \mathbb{E}[X_{n,t}]\right)\right] \\
&= \mathbb{E}\left[\sum_{i=1}^{n}\left(Y_t^{(\rho_i)} - m_1^{(\rho_i)}\right)\sum_{i=1}^{n}\left(Y_{t+s}^{(\rho_i)} - m_1^{(\rho_i)}\right)\right]
\end{aligned}, \quad s \geq 0. \quad (A.47)$$

Invoking the independence

$$\mathbb{E}\left[\left(Y_t^{(\rho_i)} - m_1^{(\rho_i)}\right)\left(Y_{t+s}^{(\rho_j)} - m_1^{(\rho_j)}\right)\right] = 0, \quad s \geq 0, \quad i \neq j, \quad (A.48)$$

we obtain

$$\begin{aligned}
\mathbb{E}\left[X_{n,t}X_{n,t+s} - \left(\mathbb{E}[X_{n,t}]\right)^2\right] &= \mathbb{E}\left[\sum_{i=1}^{n}\left(Y_t^{(\rho_i)} - m_1^{(\rho_i)}\right)\left(Y_{t+s}^{(\rho_i)} - m_1^{(\rho_i)}\right)\right] \\
&= \sum_{i=1}^{n}\mathbb{E}\left[\left(Y_t^{(\rho_i)} - m_1^{(\rho_i)}\right)\left(Y_{t+s}^{(\rho_i)} - m_1^{(\rho_i)}\right)\right] \\
&= \sum_{i=1}^{n}\mathbb{E}\left[Y_t^{(\rho_i)}Y_{t+s}^{(\rho_i)} - \left(m_1^{(\rho_i)}\right)^2\right] \\
&= \sum_{i=1}^{n}\operatorname{Var}\left[Y_t^{(\rho_i)}\right]\exp\left(-\rho_i(1-BM_1)s\right) \\
&= \sum_{i=1}^{n}\operatorname{Var}\left[Y_t^{(\rho_i)}\right]\exp\left(-\rho_i D s\right)
\end{aligned}, \quad s \geq 0. \quad (A.49)$$

Consequently, we obtain

$$\begin{aligned}
\operatorname{Cor}(s) &= \frac{\sum_{i=1}^{n}\operatorname{Var}\left[Y_t^{(\rho_i)}\right]\exp\left(-\rho_i D s\right)}{\operatorname{Var}[X_n]} \\
&= \frac{\sum_{i=1}^{n}\operatorname{Var}\left[Y_t^{(\rho_i)}\right]\exp\left(-\rho_i D s\right)}{\sum_{i=1}^{n}\operatorname{Var}\left[Y_t^{(\rho_i)}\right]} \\
&= \frac{1}{R_n}\sum_{i=1}^{n}\frac{c_i}{\rho_i}\exp\left(-\rho_i D s\right)
\end{aligned}. \quad (A.50)$$

### A.3 Statistics of infinite-dimensional supCBI process (A brief proof of Proposition 3)

The statistics of the infinite-dimensional supCBI process are derived from the results of the finite-dimensional supCBI process. The results in the previous subsection can be considered as those with

the discrete measure $\pi_n$ instead of $\pi$. Therefore, we can obtain each moment by a simple replacement

$$R_n = \int_0^{+\infty} \frac{\pi_n(\rho)}{\rho} \to R_n \equiv \int_0^{+\infty} \frac{\pi(\rho)}{\rho}, \quad (A.51)$$

due to **Corollary 1**. Therefore, we obtain (34)–(37). Similarly, we obtain

$$\mathbb{E}\left[X_t X_{t+s} - (\mathbb{E}[X_t])^2\right] = \int_0^{+\infty} \mathrm{Var}\left[Y_t^{(\rho)}\right] \exp(-\rho D s) \pi(\mathrm{d}\rho)$$
$$= \frac{AM_2}{2D^2} \int_0^{+\infty} \frac{1}{\rho} \exp(-\rho D s) \pi(\mathrm{d}\rho), \quad s \geq 0 \quad (A.52)$$

and the correlation

$$\mathrm{Cor}(s) = \frac{1}{R} \int_0^{+\infty} \frac{1}{\rho} \exp(-\rho D s) \pi(\mathrm{d}\rho), \quad s \geq 0. \quad (A.53)$$

If we assume (14), we obtain $R = \dfrac{1}{\eta(\beta-1)}$, yielding

$$\mathrm{Cor}(s) = \left(\frac{1}{1+(1-BM_1)\eta s}\right)^{\beta-1} = \left(\frac{1}{1+Us}\right)^{\beta-1}, \quad U = (1-BM_1)\eta. \quad (A.54)$$

***Remark 6*** The statistics can also be obtained from the characteristic function (19). For example, we have

$$\mathbb{E}[X_t] = -\mathrm{i} \left.\frac{\mathrm{d}C(u)}{\mathrm{d}u}\right|_{u=0} \quad (A.55)$$

and

$$\left.\frac{\mathrm{d}C(u)}{\mathrm{d}u}\right|_{u=0} \equiv \frac{\mathrm{d}}{\mathrm{d}u}\left[\exp(\mathrm{i}u\underline{X})\exp\left(A\int_0^{+\infty} \frac{\pi(\mathrm{d}\rho)}{\rho}\right)\right]_{u=0}$$
$$= \mathrm{i}\underline{X} + \frac{\mathrm{d}}{\mathrm{d}u}\left[\exp\left(AR\int_0^{+\infty}\int_0^{+\infty}(\exp(\mathrm{i}\phi_s z)-1)v(\mathrm{d}z)\mathrm{d}s\right)\right]_{u=0}$$
$$= \mathrm{i}\underline{X} + \mathrm{i}AR \cdot \int_0^{+\infty}\int_0^{+\infty} \mathrm{i}z \left.\frac{\partial \phi_s}{\partial u}\right|_{u=0} v(\mathrm{d}z)\mathrm{d}s$$
$$= \mathrm{i}\underline{X} + \mathrm{i}AM_1 R \cdot \int_0^{+\infty} \left.\frac{\partial \phi_s}{\partial u}\right|_{u=0} \mathrm{d}s$$

. (A.56)

Partially differentiating both sides of (21) with respect to $u$ yields

$$\frac{\mathrm{d}}{\mathrm{d}s}\left(\left.\frac{\partial \phi_s}{\partial u}\right|_{u=0}\right) = -D \left.\frac{\partial \phi_s}{\partial u}\right|_{u=0}, \quad s > 0, \quad \left(\left.\frac{\partial \phi_s}{\partial u}\right|_{u=0}\right)_{s=0} = 1, \quad (A.57)$$

which is solved as

$$\left.\frac{\partial \phi_s}{\partial u}\right|_{u=0} = \exp(-Ds), \quad s \geq 0. \quad (A.58)$$

Substituting (A.58) into (A.56) and then using (A.55), we obtain (34).

### A.4 Proof of Proposition 2

In this proof, positive constants $c_i$ $(i=1,2,3,4,5)$ are independent of $n$. First, we show that the right-hand side of (19) is well-defined. We assume $u \neq 0$, because the case $u = 0$ is trivial. As shown in this proof, the triple integration on the right-hand side of (19) is bounded. This holds if we have a bound

$$|f_\tau|, |g_\tau| \leq c_1 \exp(-c_2 \tau), \ g_\tau \geq -b + c_3, \ \tau > 0, \tag{A.59}$$

with constants $c_1, c_2 > 0$ and $c_3 \in (0, b)$ independent of $\rho$ but possibly depending on $u$. Indeed, we have

$$\begin{aligned}
&\left|\exp\left(i\phi_s^{(\rho)} z\right) - 1\right| \\
&= \left|\exp(-g_\tau z)\cos(f_\tau z) - 1 + i\exp(-g_\tau z)\sin(f_\tau z)\right| \\
&\leq \sqrt{\left(\exp(-g_\tau z)\cos(f_\tau z) - 1\right)^2 + \left(\exp(-g_\tau z)\sin(f_\tau z)\right)^2} \\
&\leq \left|\exp(-g_\tau z)\cos(f_\tau z) - 1\right| + \left|\exp(-g_\tau z)\sin(f_\tau z)\right| \\
&\leq \left|\exp(-g_\tau z)(\cos(f_\tau z) - 1)\right| + \left|\exp(-g_\tau z) - 1\right| + \left|\exp(-g_\tau z)\sin(f_\tau z)\right| \\
&\leq 2\exp(-g_\tau z)|f_\tau| z + \left|\exp(-g_\tau z) - 1\right| \\
&\leq 2c_1 z \exp(-c_2 \tau)\exp((b - c_3)z) + \left|\exp(-g_\tau z) - 1\right|
\end{aligned} \tag{A.60}$$

Because

$$\begin{aligned}
\left|\exp(-g_\tau z) - 1\right| &= \left(\exp(-g_\tau z) - 1\right)\chi_{\{g_\tau < 0\}} + \left(1 - \exp(-g_\tau z)\right)\chi_{\{g_\tau \geq 0\}} \\
&\leq \left(\exp\left(\min\{b - c_3, c_1 e^{-c_2 \tau}\}z\right) - 1\right)\chi_{\{g_\tau < 0\}} + g_\tau z \chi_{\{g_\tau \geq 0\}} \\
&\leq \exp\left(\min\{b - c_3, c_1 e^{-c_2 \tau}\}z\right) - 1 + c_1 z \exp(-c_2 \tau)
\end{aligned} \tag{A.61}$$

we obtain

$$\left|\exp\left(i\phi_s^{(\rho)} z\right) - 1\right| \leq \exp\left(\min\{b - c_3, c_1 e^{-c_2 \tau}\}z\right) - 1 + 3c_1 z \exp(-c_2 \tau). \tag{A.62}$$

Now, note that

$$\begin{aligned}
&\left|\int_0^{+\infty}\int_0^{+\infty}\int_0^{+\infty}\left(\exp\left(i\phi_s^{(\rho)} z\right) - 1\right)v(dz)ds\pi(d\rho)\right| \\
&= \left|\int_0^{+\infty}\int_0^{+\infty}\int_0^{+\infty}\left(\exp(i\phi_\tau z) - 1\right)v(dz)d\tau \frac{\pi(d\rho)}{\rho}\right| \\
&\leq \int_0^{+\infty}\int_0^{+\infty}\int_0^{+\infty}\left|\exp(i\phi_\tau z) - 1\right|v(dz)d\tau \frac{\pi(d\rho)}{\rho} \\
&= \int_0^{+\infty}\frac{\pi(d\rho)}{\rho} \cdot \int_0^{+\infty}\int_0^{+\infty}\left|\exp(i\phi_\tau z) - 1\right|v(dz)d\tau
\end{aligned} \tag{A.63}$$

On the last term of (A.62), we have

$$\int_0^{+\infty}\int_0^{+\infty}\exp(-c_2 \tau)zv(dz)d\tau = \int_0^{+\infty}\exp(-c_2 \tau)d\tau \cdot \int_0^{+\infty}zv(dz)$$
$$= \frac{1}{c_2}\int_0^{+\infty}zv(dz) \tag{A.64}$$
$$< +\infty$$

On the remaining terms of (A.62), if $b - c_3 \geq c_1$, we have

$$\int_0^{+\infty}\int_0^{+\infty}\left\{\exp\left(\min\left\{b-c_3,c_1e^{-c_2\tau}\right\}z\right)-1\right\}v(\mathrm{d}z)\mathrm{d}\tau$$
$$=\int_0^{+\infty}\int_0^{+\infty}\left\{\exp\left(c_1e^{-c_2\tau}z\right)-1\right\}v(\mathrm{d}z)\mathrm{d}\tau$$
$$\leq \int_0^{+\infty}\int_0^{+\infty}c_1e^{-c_2\tau}zv(\mathrm{d}z)\mathrm{d}\tau \qquad\qquad (A.65)$$
$$=\frac{c_1}{c_2}\int_0^{+\infty}zv(\mathrm{d}z)$$
$$<+\infty$$

If $b-c_3<c_1$, for with some constant $c_4>0$ we have

$$\int_0^{+\infty}\int_0^{+\infty}\left\{\exp\left(\min\left\{b-c_3,c_1e^{-c_2\tau}\right\}z\right)-1\right\}v(\mathrm{d}z)\mathrm{d}\tau$$
$$=\int_0^{c_4}\int_0^{+\infty}\left\{\exp\left(\min\left\{b-c_3,c_1e^{-c_2\tau}\right\}z\right)-1\right\}v(\mathrm{d}z)\mathrm{d}\tau$$
$$+\int_{c_4}^{+\infty}\int_0^{+\infty}\left\{\exp\left(\min\left\{b-c_3,c_1e^{-c_2\tau}\right\}z\right)-1\right\}v(\mathrm{d}z)\mathrm{d}\tau \qquad (A.66)$$
$$=+\int_0^{c_4}\int_0^{+\infty}\left\{\exp\left((b-c_3)z\right)-1\right\}v(\mathrm{d}z)\mathrm{d}\tau+\int_{c_4}^{+\infty}\int_0^{+\infty}\left\{\exp\left(c_1e^{-c_2\tau}z\right)-1\right\}v(\mathrm{d}z)\mathrm{d}\tau$$
$$\leq c_4\int_0^{+\infty}\left\{\exp\left((b-c_3)z\right)-1\right\}v(\mathrm{d}z)+\int_0^{+\infty}\int_0^{+\infty}c_1e^{-c_2\tau}zv(\mathrm{d}z)\mathrm{d}\tau$$
$$<+\infty$$

By (A.65) and $\int_0^{+\infty}\left\{\exp\left((b-c_3)z\right)-1\right\}v(\mathrm{d}z)<+\infty$ due to $c_3\in(0,b)$, we obtain the boundedness

$$\left|\int_0^{+\infty}\int_0^{+\infty}\int_0^{+\infty}\left(\exp\left(\mathrm{i}\phi_s^{(\rho)}z\right)-1\right)v(\mathrm{d}z)\mathrm{d}s\pi(\mathrm{d}\rho)\right|<+\infty. \qquad (A.67)$$

Consequently, it is sufficient to verify (A.59) to well-define (19). From Proposition 3.10 of Jin et al. (2020), smooth solutions to (23)–(24) satisfy the first bound $|f_\tau|,|g_\tau|\leq c_1\exp(-c_2\tau)$ ($\tau\geq 0$) given in (A.59) and exist globally in time. Theorem 2.3 of the same literature guarantees the uniqueness of the solutions to (23)–(24). The second bound $g_\tau\geq -b+c_3$ is a byproduct of the first bound. Indeed, if it is not true, $g_\tau\leq -b$ at some $\tau>0$ at which the right-hand side of (23) diverges unless $f_\tau=0$. However, this does not occur at finite $\tau>0$ by $u\neq 0$ and the smoothness of $f_\tau$ is a contradiction. Therefore, we have (A.67), yielding that (19) is well-defined.

We analyze the convergence of $C_n$ to $C$. To this end, we denote the main branch of a complex variable by "Log." We evaluate the difference

$$D(u)=\left|\mathrm{Log}\,C_n(u)-\mathrm{Log}\,C(u)\right|,\ u\in\mathbb{R}. \qquad (A.68)$$

For $u\in\mathbb{R}$, we have

$$
\begin{aligned}
D_n(u) &= A\left|\begin{array}{l}\int_0^{+\infty}\int_0^{+\infty}\int_0^{+\infty}(\exp(i\phi_\tau z)-1)v(\mathrm{d}z)\mathrm{d}\tau\dfrac{\pi_n(\mathrm{d}\rho)}{\rho}\\ -\int_0^{+\infty}\int_0^{+\infty}\int_0^{+\infty}(\exp(i\phi_\tau z)-1)v(\mathrm{d}z)\mathrm{d}\tau\dfrac{\pi(\mathrm{d}\rho)}{\rho}\end{array}\right|\\
&= A\left|\int_0^{+\infty}\int_0^{+\infty}\left((\exp(i\phi_\tau z)-1)\right)v(\mathrm{d}z)\mathrm{d}\tau\left(\int_0^{+\infty}\dfrac{\pi_n(\mathrm{d}\rho)}{\rho}-\int_0^{+\infty}\dfrac{\pi(\mathrm{d}\rho)}{\rho}\right)\right|\\
&\leq A\left|\int_0^{+\infty}\int_0^{+\infty}\left((\exp(i\phi_\tau z)-1)\right)v(\mathrm{d}z)\mathrm{d}\tau\right|\left|\int_0^{+\infty}\dfrac{\pi_n(\mathrm{d}\rho)}{\rho}-\int_0^{+\infty}\dfrac{\pi(\mathrm{d}\rho)}{\rho}\right|
\end{aligned}
\qquad (\text{A.69})
$$

Then, we as follows:

$$
D_n(u)\leq c_5\left|\int_0^{+\infty}\dfrac{\pi_n(\mathrm{d}\rho)}{\rho}-\int_0^{+\infty}\dfrac{\pi(\mathrm{d}\rho)}{\rho}\right|,\quad u\in\mathbb{R}, \qquad (\text{A.70})
$$

with some $c_5>0$, showing that $D_n(u)$ is essentially ruled by

$$
d_n\equiv\left|\int_0^{+\infty}\dfrac{\pi_n(\mathrm{d}\rho)}{\rho}-\int_0^{+\infty}\dfrac{\pi(\mathrm{d}\rho)}{\rho}\right|. \qquad (\text{A.71})
$$

Thus, we focus on the analysis of $d_n$. We show that by choosing the specified sequences $\{c_i,\rho_i\}_{1\leq i\leq n}$, it follows that $d_n\to +0$ as $n\to +\infty$ and hence the consistency of the finite-dimensional supCBI process.

With (25)–(26), we can evaluate $d_n$ as follows

$$
\begin{aligned}
d_n &= \left|\sum_{i=1}^n\int_{x_{i-1}}^{x_i}\left(\dfrac{\pi_n(\mathrm{d}\rho)}{\rho}-\dfrac{\pi(\mathrm{d}\rho)}{\rho}\right)\right|\\
&= \left|\sum_{i=1}^n\left(\dfrac{c_i}{\rho_i}-\int_{x_{i-1}}^{x_i}\dfrac{\pi(\mathrm{d}\rho)}{\rho}\right)\right|\\
&= \left|\sum_{i=1}^n\int_{x_{i-1}}^{x_i}\left(\dfrac{\pi(\mathrm{d}\rho)}{\rho_i}-\dfrac{\pi(\mathrm{d}\rho)}{\rho}\right)\right|\\
&= \left|\sum_{i=1}^n\int_{x_{i-1}}^{x_i}\left(\dfrac{1}{\rho_i}-\dfrac{1}{\rho}\right)\pi(\mathrm{d}\rho)\right|\\
&= \left|\sum_{i=1}^n\int_{x_{i-1}}^{x_i}\dfrac{\rho_i-\rho}{\rho_i\rho}\pi(\mathrm{d}\rho)\right|\\
&\leq \sum_{i=1}^{n-1}\dfrac{1}{\rho_i}\int_{x_{i-1}}^{x_i}\dfrac{|\rho-\rho_i|}{\rho}\pi(\mathrm{d}\rho)+\dfrac{1}{x_{n-1}}\int_{x_{n-1}}^{+\infty}\dfrac{\rho-x_{n-1}}{\rho}\pi(\mathrm{d}\rho)\\
&\leq \sum_{i=1}^{n-1}\dfrac{1}{\rho_i}\int_{x_{i-1}}^{x_i}\dfrac{|\rho-\rho_i|}{\rho}\pi(\mathrm{d}\rho)+\dfrac{1}{x_{n-1}}\int_{x_{n-1}}^{+\infty}\pi(\mathrm{d}\rho)
\end{aligned}
\qquad (\text{A.72})
$$

We have

$$
\begin{aligned}
\sum_{i=1}^{n-1}\dfrac{1}{\rho_i}\int_{x_{i-1}}^{x_i}\dfrac{|\rho-\rho_i|}{\rho}\pi(\mathrm{d}\rho) &= \sum_{i=1}^{n-1}\dfrac{2}{x_i+x_{i-1}}\int_{x_{i-1}}^{x_i}\dfrac{|\rho-(x_i+x_{i-1})/2|}{\rho}\pi(\mathrm{d}\rho)\\
&\leq \sum_{i=1}^{n-1}\dfrac{2(x_i-x_{i-1})}{x_i+x_{i-1}}\int_{x_{i-1}}^{x_i}\dfrac{1}{\rho}\pi(\mathrm{d}\rho)
\end{aligned}
\qquad (\text{A.73})
$$

since $|\rho-(x_i+x_{i-1})/2|\leq x_i-x_{i-1}$ for $\rho\in[x_{i-1},x_i]$. Consequently, we obtain

$$d_n \leq \sum_{i=1}^{n-1}\frac{2(x_i-x_{i-1})}{x_i+x_{i-1}}\int_{x_{i-1}}^{x_i}\frac{1}{\rho}\pi(\mathrm{d}\rho)+\frac{1}{x_{n-1}}\int_{x_{n-1}}^{+\infty}\pi(\mathrm{d}\rho). \tag{A.74}$$

Then, we obtain the estimate

$$\begin{aligned}d_n &\leq \sum_{i=1}^{n-1}\frac{2\left(\overline{C}in^{-\gamma}-\overline{C}(i-1)n^{-\gamma}\right)}{\overline{C}in^{-\gamma}+\overline{C}(i-1)n^{-\gamma}}\int_{x_{i-1}}^{x_i}\frac{1}{\rho}\pi(\mathrm{d}\rho)+\frac{1}{x_{n-1}}\int_{x_{n-1}}^{+\infty}\pi(\mathrm{d}\rho)\\ &= \sum_{i=1}^{n-1}\frac{1}{i-1/2}\int_{x_{i-1}}^{x_i}\frac{1}{\rho}\pi(\mathrm{d}\rho)+\frac{1}{x_{n-1}}\int_{x_{n-1}}^{+\infty}\pi(\mathrm{d}\rho)\\ &= C_0\sum_{i=1}^{n-1}\frac{1}{i-1/2}(x_i-x_{i-1})^\delta+\frac{1}{x_{n-1}}\int_{x_{n-1}}^{+\infty}\pi(\mathrm{d}\rho)\\ &\leq C_0\overline{C}^\delta\frac{1}{n^{\gamma\delta}}\sum_{i=1}^{n-1}\frac{1}{i-1/2}+\frac{1}{x_{n-1}}\int_{x_{n-1}}^{+\infty}\pi(\mathrm{d}\rho)\end{aligned} \tag{A.75}$$

Thus, we have $d_n\to+0$ as $n\to+\infty$ because $d_n\geq 0$, the second term on the right-hand side of (A.75) converges to 0 as $n\to+\infty$, and the first term is evaluated as $\lim_{n\to+\infty}\frac{1}{n^{\gamma\delta}}\sum_{i=1}^{n-1}\frac{1}{i-1/2}=\lim_{n\to+\infty}\frac{1}{n^{\gamma\delta}}\ln n=0$, which completes the proof.

□

**Remark 7** The choice of each $x_i$ ($1\leq i\leq n-1$) can be more flexible if necessary provided there exist constants $C_0>0$ and $\delta\in(0,1]$ independent of $n$ such that $\int_{x_{i-1}}^{x_i}\frac{1}{\rho}\pi(\mathrm{d}\rho)\leq C_0(x_i-x_{i-1})^\delta$ ($1\leq i\leq n-1$).